\documentclass[letterpaper,11pt]{article}
\pdfoutput=1

\usepackage{jheppub}
\usepackage{graphicx}
\usepackage{amsmath}
\usepackage{amssymb}
\usepackage{mathrsfs}
\usepackage{bm}
\usepackage{braket}
\usepackage{slashed}
\usepackage{xspace}
\usepackage[normalem]{ulem}
\usepackage[utf8]{inputenc}
\usepackage{multirow}
\usepackage{courier}

\newcommand{\eq}[1]{eq.~\eqref{eq:#1}}
\newcommand{\eqs}[2]{eqs.~\eqref{eq:#1} and \eqref{eq:#2}}
\renewcommand{\sec}[1]{sec.~\ref{sec:#1}}
\newcommand{\secs}[2]{secs.~\ref{sec:#1} and \ref{sec:#2}}
\newcommand{\app}[1]{app.~\ref{sec:#1}}

\newcommand{\fig}[1]{fig.~\ref{fig:#1}}

\newcommand{\ord}[1]{{\mathcal O}(#1)}
\newcommand{\nn}{\nonumber}
\newcommand{\vecb}[1]{\mbox{\boldmath $#1$}}

\newcommand{\df}{\mathrm{d}}
\newcommand{\img}{\mathrm{i}}
\newcommand{\sdt}{\!\cdot\!}

\newcommand{\al}{\alpha}

\newcommand{\ga}{\gamma}
\newcommand{\Ga}{\Gamma}
\newcommand{\de}{\delta}

\newcommand{\ve}{\varepsilon}
\newcommand{\la}{\lambda}
\newcommand{\si}{\sigma}

\newcommand{\thh}{\theta}
\newcommand{\Thh}{\Theta}
\newcommand{\ze}{\zeta}

\newcommand{\cL}{{\mathcal L}}

\newcommand{\cR}{{\mathcal R}}
\newcommand{\cO}{{\mathcal O}}

\newcommand{\bnslash}{\bar{n}\!\!\!\slash}
\newcommand{\bn}{{\bar{n}}}

\newcommand{\sandwich}[3]{\left< #1 \right | #2 \left | #3 \right >}

\def\({\left(}
\def\[{\left[}
\def\){\right)}
\def\]{\right]}

\newcommand{\WTA}{\mathrm{WTA}}
\newcommand{\SJA}{\mathrm{SJA}}
\newcommand{\case}{\mathrm{case}}
\newcommand{\cut}{\mathrm{cut}}

\newcommand{\zero}{{[0]}}
\newcommand{\one}{{[1]}}
\newcommand{\two}{{[2]}}
\newcommand{\three}{{[3]}}

\newcommand{\Eventtwo}{\texttt{Event2}\xspace}
\newcommand{\Artemide}{\texttt{arTeMiDe}\xspace}

\newcommand{\LzeroQT}{\mathcal{L}_0(q_T,\mu)}
\newcommand{\LoneQT}{\mathcal{L}_1(q_T,\mu)}
\newcommand{\LzeroQTR}{\mathcal{L}_0^\cut\Big(q_T,\frac{E\cR}{2}\Big)}
\newcommand{\LoneQTR}{\mathcal{L}_1^\cut\Big(q_T,\frac{E\cR}{2}\Big)}
\newcommand{\LzeroQTZ}{\mathcal{L}_0^\cut\big(q_T,E\cR(1-z)\big)}
\newcommand{\thetaQT}{\thh\Big(\frac{E\cR}{2}-q_T\Big)}

\allowdisplaybreaks[2]

\preprint{\vbox{\hbox{Nikhef 2019-009}}}

\title{Transverse momentum dependent distributions in $\mathbf{e^+e^-}$ and semi-inclusive deep-inelastic scattering using jets}

\author[a]{Daniel Gutierrez-Reyes,}
\author[a]{Ignazio Scimemi,}
\author[b,c]{Wouter J.~Waalewijn,}
\author[b,c]{Lorenzo Zoppi}

\affiliation[a]{Departamento de F\'isica Te\'orica, Universidad Complutense de Madrid (UCM) and IPARCOS, E-28040 Madrid, Spain}
\affiliation[b]{Institute for Theoretical Physics Amsterdam and Delta Institute for Theoretical Physics, University of Amsterdam, Science Park 904, 1098 XH Amsterdam, The Netherlands}
\affiliation[c]{Nikhef, Theory Group, Science Park 105, 1098 XG, Amsterdam, The Netherlands}

\abstract{The extraction of transverse momentum dependent distributions (TMDs) in semi-inclusive deep inelastic scattering (SIDIS) is complicated by the presence of both initial- and final-state nonperturbative physics. We recently proposed measuring jets (instead of hadrons) as a solution, showing that for the Winner-Take-All jet axis the same factorization formulae valid for hadrons applied to jets of arbitrary size. This amounts to simply replacing TMD fragmentation functions  by our TMD jet functions. In this paper we present the calculation of these jet functions at one loop. We obtain phenomenological results for $e^+e^- \to$~dijet (Belle II, LEP) and SIDIS (HERA, EIC) with a jet, building on the \Artemide code. Surprisingly, we find that the limit of large jet radius describes the full $R$ results extremely well, and we extract the two-loop jet function in this limit using \Eventtwo, allowing us to achieve N$^3$LL accuracy. We demonstrate the perturbative convergence of our predictions and explore the kinematic dependence of the cross section. Finally, we investigate the sensitivity to nonperturbative physics, demonstrating that jets are a promising probe of proton structure.}
\begin{document}
\maketitle

%%%%%%%%%%%%%%%%%%%%%%%%%%%%%%%%%%%%%%%%%%%%%%%%%%%%%%%%%%%%%%%%%%%%%%%%%%%%%%%%
\section{Introduction}
\label{sec:intro}
%%%%%%%%%%%%%%%%%%%%%%%%%%%%%%%%%%%%%%%%%%%%%%%%%%%%%%%%%%%%%%%%%%%%%%%%%%%%%%%%

Since the early days of the parton model, the structure of the proton has been a major focus of the nuclear and particle physics communities. In addition to being of intrinsic interest, it is of direct relevance for describing the initial state at hadron colliders such as the LHC, and therefore important in the search for new short-distance physics. The essential theoretical ingredient is factorization, which allows one to separate the cross section into a hard scattering, that can be calculated in perturbation theory, and process-independent parton distribution functions (PDFs). The PDFs parametrize the proton structure, describing the momentum fraction of partons in the proton along the direction of motion. 

We will focus on transverse momentum dependent PDFs, where in addition the transverse momentum of partons in the proton is probed. Since a transverse momentum measurement can also be thought of as the measurement of an angle, it is natural that TMD factorization theorems generically involve two TMD distributions. Traditionally, the relative transverse momentum of two hadrons in $e^+e^-$, the transverse momentum of a hadron semi-inclusive deep-inelastic scattering (SIDIS), $ep \to ehX$, and the transverse momentum of a $\ga^*/Z$ boson in $pp$ collisions have been considered. 

We recently proposed replacing individual final-state hadrons by jets in the above measurements~\cite{Gutierrez-Reyes:2018qez}. Jets are collimated sprays of hadrons, that appear in high-energy collisions because of the collinear singularity of quantum chromodynamics (QCD). Practically they are identified by clustering particles according to a specified algorithm. 
On the theoretical side, we demonstrated that one can simply replace the TMD fragmentation functions entering factorization theorems with our TMD jet functions, for which the use of the Winner-Take-All (WTA) recombination scheme~\cite{Bertolini:2013iqa} played a key role. 
The advantage of our approach is that such functions are perturbatively calculable, thus removing an important source of uncertainty. Specifically, the intrinsically nonperturbative distribution of the momentum fraction of individual hadrons is removed by using jets. 

In the context of SIDIS experiments, replacing the nonperturbative TMD fragmentation functions with calculable jet functions would allow one to increase the sensitivity to initial-state nonperturbative physics. It will be interesting to see whether this can be investigated with existing HERA data, and exciting to explore at the electron-ion collider (EIC), which will enable the extraction of PDFs with unmatched precision, with SIDIS experiments playing an important role~\cite{Aschenauer:2019kzf}. Of course, for small transverse momenta, the jet functions themselves will also receive nonperturbative corrections. However, this can be addressed by exploiting the universality of the nonperturbative structure of the TMD jet function, with $e^+e^- \rightarrow$ dijet providing a useful testing ground. Explicitly, data from $e^+e^-$ collisions could be used to fit a model for nonperturbative corrections to the jet function to be later applied to SIDIS. 

A number of other jet observables that account for transverse momentum dependence have recently been considered. The main focus has been on the transverse momentum of hadrons fragmenting in jets, in both inclusive~\cite{Bain:2016rrv} and semi-inclusive~\cite{Kang:2017yde,Kang:2017btw} processes. In the same context, refs.~\cite{Makris:2017arq,Makris:2018npl} used soft-drop jet grooming~\cite{Larkoski:2014wba} to reduce sensitivity to soft radiation within the jet. These studies consider the transverse momentum with respect to the standard jet axis (SJA); instead, as an alternative way to reduce sensitivity to soft radiation, refs.~\cite{Neill:2016vbi,Neill:2018wtk} performed a similar analysis for the transverse momentum with respect to the Winner-Take-All (WTA) axis.
The transverse momentum of the jet itself was also recently considered in photon + jet production~\cite{Buffing:2018ggv} and lepton-jet correlation in deep-inelastic scattering~\cite{Liu:2018trl}.
 
Besides showing a full derivation of the results presented in ref.~\cite{Gutierrez-Reyes:2018qez}, the main purpose of this paper is performing a numerical analysis of $e^+ e^- \rightarrow$ dijet and semi-inclusive deep-inelastic scattering (SIDIS) using  \Artemide~\cite{Scimemi:2017etj,Scimemi:2018xaf}, to study the phenomenology of TMDs with jets.

%%%
\begin{figure*}[tb]
  \centering
  \includegraphics[width=0.75\textwidth]{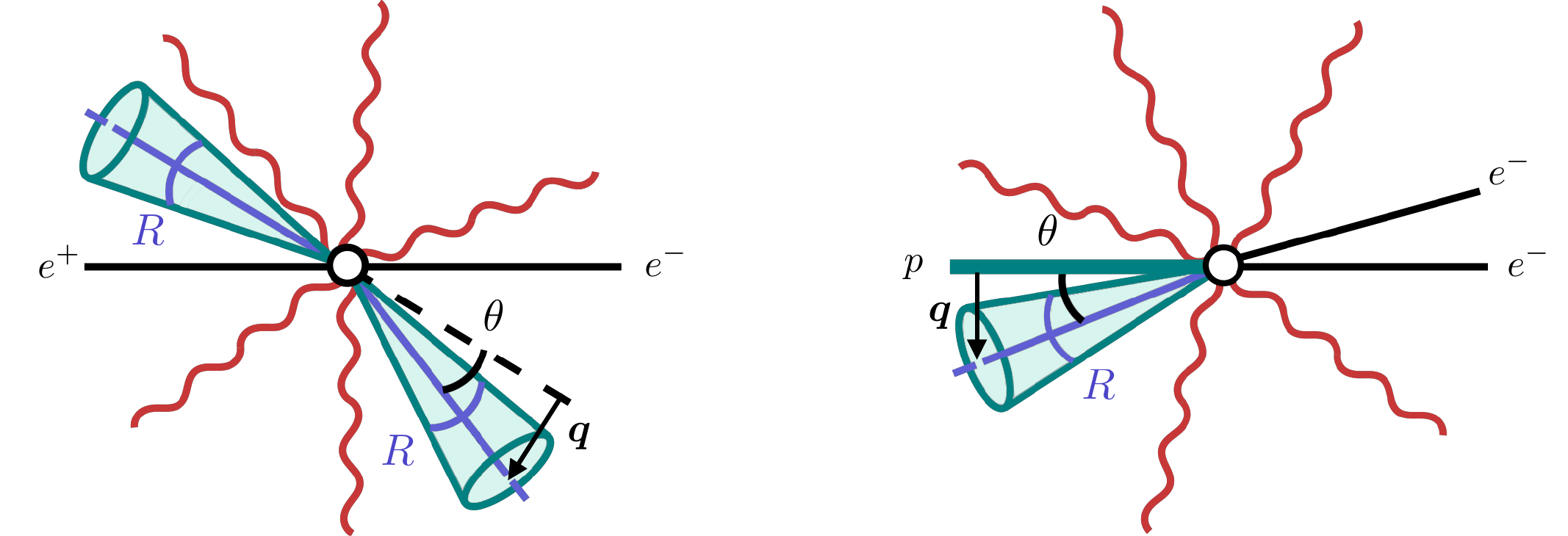}%
  \caption{Geometry of the event for $e^+e^- \rightarrow$ dijet (left) and SIDIS (right). The horizontal direction represents the beam axis. For dijets the relevant quantities $\vecb{q}$ and $\theta$ are the transverse momentum decorrelation and angular decorrelation of the system, defined with respect to the relative orientation of the two jets. We consider almost back-to-back jets, $\theta \ll 1$, and study different hierarchies between $\theta$ and the jet radius $R$. In SIDIS $\vecb{q}$ represents the transverse mometum of the jet, and the corresponding angle is measured with respect to the beam axis. We work in the Breit frame, where the jet recoils almost in the direction of the incoming proton, $\theta \ll 1$.}
\label{fig:event_sketch}
\end{figure*}
%%%

In the case of $e^+ e^- \rightarrow $ dijet, the main physical quantity we consider is the transverse momentum decorrelation. It is defined as\footnote{In this paper, we reserve bold font for denoting transverse two-vector quantities.}
%%%
\begin{align} \label{eq:def:decorrelation}
 \qquad\qquad \vecb{q} = \frac{\vecb{p}_{1}}{z_1} + \frac{\vecb{p}_{2}}{z_2}\, , \qquad \qquad (e^+e^- \rightarrow \mbox{dijet})
\end{align}
%%%
where $\vecb{p}_{i}$ are the jet transverse momenta measured with respect to a common direction and $z_i = 2E_i/\sqrt{s}$ are their energy fractions, $\sqrt{s}$ is the center-of-mass energy of the collision. 
Since factorization requires a small transverse momentum decorrelation, we will always assume
%%%
\begin{align} \label{eq:qTsmall}
  q_T \equiv | \vecb{q} | \ll \frac{\sqrt{s}}{{2}}\, .
\end{align}
%%%
A related quantity is the angular decorrelation, shown in the left panel of \fig{event_sketch},
%%%
\begin{align} \label{eq:def:angular_decorrelation}
  \theta = \arctan \Big(\frac{2q_T}{\sqrt{s}}\Big) \approx \frac{2q_T}{\sqrt{s}}\, ,
\end{align}
%%%
where the final expression exploits \eq{qTsmall}. This makes it explicit that we consider configurations where jets are almost back to back.\footnote{Another interesting small-angle configuration occurs for two jets moving in almost the same direction, which we do not study in this paper.} The angular decorrelation is similar to the azimuthal decorrelation in hadronic collisions, calculated at next-to-leading logarithmic accuracy in refs.~\cite{Banfi:2008qs,Sun:2014gfa,Sun:2015doa,Chen:2016cof}. 

In principle, the definitions in \eqs{def:decorrelation}{def:angular_decorrelation} depend on the choice of axis with respect to which the jet transverse momenta are measured. However, differences induced by this choice are suppressed by powers of $q_T^2/s$. Of course, the definition \emph{is} sensitive to the details of the jet algorithm: our default throughout the paper will be the WTA axis with anti-$k_T$ \cite{Cacciari:2008gp}, but we will also consider  the SJA and other clustering algorithms of the $k_T$ family. We will also explore the dependence on the jet radius $R$.

In SIDIS, shown in the right panel of \fig{event_sketch}, we choose to work in the Breit frame and define the transverse momentum as
%%%
\begin{align}
  \qquad \qquad \vecb{q} = \frac{\vecb{P}_{J}}{z} + \vecb{q}_{\rm in}\, ,\qquad \qquad \mbox{(SIDIS)}
\end{align}
%%%
where $\vecb{P}_J$ is the transverse momentum of the jet with respect to the beam axis, $\vecb{q}_{\rm{in}}$ is the transverse momentum of the initial-state quark in the proton, and $z=2E_J/Q$ is the jet energy normalized to (minus) the virtuality of the photon $Q^2$. In analogy with \eq{def:angular_decorrelation} we define a corresponding angle $\theta$ and require 
%%%
\begin{align}
  \theta = \arctan\Big(\frac{2q_T}{Q}\Big) \simeq \frac{2q_T}{Q} \ll 1\, .
\end{align}
%%%
We use the same symbols $q_T$ and $\theta$ for analog quantities in different processes since they play the same role in factorization formulae, and their meaning should be clear from the context. 

To summarize our main findings: when using the WTA axis, the same factorization formulae valid for hadrons hold for jets, independently of the hierarchy between the angle $\theta$ and the jet radius parameter $R$.  
Because  the factorization theorem ensures that hadronization effects in the jets are universal,  they can be estimated in $e^+e^-$ and  then used in the analysis of SIDIS experiments.  We anticipate  that  the main nonperturbative effects come from the evolution factor. These effects are universal (i.e.~the same in $e^+e^-$, SIDIS, and Drell-Yan experiments and  independent of the polarization of the hadrons)
and their estimation is one of the major goals of  TMD analyses. 
In this context we note the vital role played by the $\zeta$-prescription~\cite{Scimemi:2018xaf}, which ensures that the nonperturbative contribution to the evolution factor (that is responsible for the resummation) is uncorrelated with other nonperturbative effects.

Another observation has lead us to focus on the large radius regime of the jets. In fact, at one-loop order we notice that our jet function is well described by its large-$R$ limit. In this limit the jet functions  simplify considerably, and are determined by renormalization group evolution (RGE) up to a constant. We exploit this fact to numerically extract the two-loop, large-radius jet function from \Eventtwo and push the accuracy of the calculation to N$^3$LL in this case. Surprisingly, the validity of this regime extends down to fairly small values of the jet radius, allowing us to get precise results across the whole range in transverse momentum. This brings the perturbative precision of TMDs with jets on par with TMDs with final-state hadrons.

The paper is structured as follows: In \sec{fact} we discuss the factorization formulae, considering different hierarchies between $R$ and $\theta$, illustrated in \fig{event_sketch}. We present expressions for both the transverse momentum decorrelation in $e^+e^- \to$ dijet, as well as the transverse momentum of the jet in SIDIS. In addition to our default choice of using the WTA axis, we also consider the standard jet axis (SJA), for which we show that the factorization is significantly more complicated when $\theta \ll R$. In \sec{jet} we explicitly compute the quark jet functions at one-loop order, performing the calculation in both transverse-momentum and impact-parameter space. The renormalization and resummation is discussed in \sec{resum}, and the two-loop jet function for $\theta \ll R$ is extracted from \Eventtwo in \sec{jet_two}.  In \sec{results} we present our numerical results for $e^+e^-\to$ dijet and SIDIS, and we conclude in \sec{conc}. A summary of our conventions and perturbative ingredients are collected in the appendix.

%%%%%%%%%%%%%%%%%%%%%%%%%%%%%%%%%%%%%%%%%%%%%%%%%%%%%%%%%%%%%%%%%%%%%%%%%%%%%%%%
\section{Factorization of the cross section and definition of the jet functions}
\label{sec:fact}
%%%%%%%%%%%%%%%%%%%%%%%%%%%%%%%%%%%%%%%%%%%%%%%%%%%%%%%%%%%%%%%%%%%%%%%%%%%%%%%%

The factorization of the cross section depends on  the quantity
%%%
\begin{align}
\label{eq:jetsize}
{\cal R}\equiv 2 \tan \frac{R}{2}
\,.\end{align}
%%%
For small values, $\cR$ is just the jet radius parameter $R$, but in general the parameter $\cR$ allows us to capture some power corrections. 
In the following we will use $\cR$ when considering transverse momenta, while we use $R$ when considering angles.
In this section we review the factorization formulae of ref.~\cite{Gutierrez-Reyes:2018qez} for all possible hierarchies, while in the remainder of this paper we concentrate on the ones that play a role in our phenomenological results. We start here by introducing the jet function, which is the main new ingredient of our analysis, providing its definition and briefly discussing its renormalization.

Our factorization analysis is carried out using Soft-Collinear Effective Theory (SCET)~\cite{Bauer:2000ew, Bauer:2000yr, Bauer:2001ct, Bauer:2001yt}, in which the jets are described by collinear modes and the radiation outside the jets is described by a soft mode. The typical momentum scaling of these modes are summarized in table \ref{tab:modes}, in terms of light-cone coordinates
%%%
\begin{align}
 p^\mu =  (p^-, p^+, \vecb{p}) = p^- \frac{n^\mu}{2} + p^+ \frac{\bn^\mu}{2} + \vecb{p}^\mu
\,.\end{align}
%%%
Here $n^\mu$ and $\bn^\mu$ are light-like vectors along the directions of the jets, with $n \cdot \bn = 2$.

The jet function, that enters the factorization theorem for $\theta \sim R$, is written in $\vecb{b}$-space as the following collinear matrix element 
%%%
\begin{align} \label{eq:Jq_def}
   J_q(z, \vecb{b},  E {\cal R}) &=\frac{z}{2N_c}\text{Tr}\Big[\frac{\bnslash}{2} 
\bra{0}\big[\delta \big(
2E/z-\bar n \sdt P\big) 
e^{\img \textit{\textbf{b}}\cdot\textit{\textbf{{P}}}}  \chi_n(0) \Big]
\sum_X |J_{\text{alg},R} X\rangle \langle J_{\text{alg},R}X|\bar\chi_n(0)\ket{0}
.\end{align}
%%%
Here, $z$ is the light-cone momentum fraction of the jet with respect to the initiating quark, $E$ is the energy of the initiating quark, and $P$ is the momentum operator. The trace in \eq{Jq_def} is over Dirac indices, and $\chi_n(y) = W_n^\dagger(y) \xi_n(y)$, where $\xi_n$ is the collinear quark field in the light-like direction $n^\mu$ and $W_n$ is a collinear Wilson line, ensuring collinear gauge invariance.
The subscript \emph{alg} serves as a reminder that the jet function depends on the clustering algorithm, which works as follows: as long as at least one pair of particles exists whose angular distance is smaller than $R$, the two particles with the smallest distance measure are selected and merged. The rule to merge two particles of four-momenta $p_1, p_2$ into a new ``particle" with momentum $p_{(12)}$ reads
%%%
\begin{align} \label{eq:mergingPrescription}
  \mbox{SJA}: E_{(12)} &= E_1 + E_2,\quad \vec{p}_{(12)} = \vec{p}_1 + \vec{p}_2\,, \nonumber \\
  \mbox{WTA}: E_{(12)} &= E_1 + E_2,\quad \vec{p}_{(12)} = 
  E_{(12)}\Big[\frac{\vec{p}_1}{|\vec{p}_1|}\theta(E_1-E_2) +
  \frac{\vec{p}_2}{|\vec{p}_2|}\theta(E_2-E_1)\Big]\,,
\end{align}
%%%
i.e.~with the SJA the two four-momenta are added, while with the WTA the new pair is massless by definition, and its direction coincides with the one of the most energetic particle. The algorithm stops when the angular separation between each pair of remaining particles exceeds $R$, in which case these ``particles" are considered the final jets.

Gluon-initiated jets do not enter for $e^+e^-$ and SIDIS, but we give the corresponding definition for completeness, 
%%%
\begin{align}
J_g(z,\vecb{b},E\mathcal{R}) &= \frac{zE}{N_c^2-1} \bra{0} 
\big[\delta \big(
2E/z-\bar n \cdot P\big) 
e^{\img \textit{\textbf{b}}\cdot\textit{\textbf{{P}}}} 
\,\mathcal{B}_{n\perp\mu}(0) \big] \ket{J_{\text{alg},R}X}\bra{J_{\text{alg},R}X}\mathcal{B}_{n\perp\mu}(0)\ket{0}
,\end{align}
%%%
where 
%%%
\begin{align}
\mathcal{B}_{n\perp\mu} = \frac{1}{\bar n \cdot {\cal P}}\, \img \bar n _\alpha g_{\perp \mu\beta}W_n^\dagger F_n^{\alpha\beta}W_n \end{align}
%%%
is the collinear gluon field, with $F_n^{\alpha\beta}$ the collinear field strength tensor.
We will also perform the calculation in momentum space, which simply involves replacing 
%%%
\begin{align}
   e^{\img \textit{\textbf{b}}\cdot\textit{\textbf{{P}}}}   \quad \to \quad
   \int\! \frac{\df^2 \vecb{b}}{(2\pi)^2}\, e^{\img \textit{\textbf{b}}\cdot(\textit{\textbf{P}}-\textit{\textbf{q}})} 
   = \de^2(\vecb{q} - \vecb{P})
\,.\end{align}
%%%

The above definitions are for the bare jet functions, as indicated by the absence of renormalization scales.
A perturbative calculation shows that both ultraviolet (UV) and rapidity divergences affect these distributions, so that one should consider the renormalized quantities
%%%
\begin{align}
J_q(z,\vecb{b},E\mathcal{R},\mu,\zeta) =Z_q(\zeta,\mu) R_q(\zeta,\mu) J_q(z,\vecb{b},E\mathcal{R}) 
\label{eq:ZR}
\end{align}
%%%
and  similarly for $J_g$.  Here $Z_q$ is the UV renormalization factor, $R_q$ is the rapidity renormalization factor, and rapidity divergences are removed first, as in ref.~\cite{Echevarria:2016scs}. 
A key observation is that these renormalization factors are the same as in the case of TMDs, as we discuss in sec.~\ref{sec:resum}.

  \begin{table}
   \centering
   \begin{tabular}{l|ccccc}
     \hline \hline
      Mode & $ {R} \ll \theta \ll 1$  & $\theta \sim {R} \ll 1$ & $\theta \ll {R}$ (WTA) & $\theta \ll {R}\ll 1$ (SJA) & $\theta \ll {R}\sim 1$ (SJA) \\ \hline
      hard & (1,1,1) & (1,1,1) & (1,1,1) & (1,1,1) & (1,1,1) \\
     $n$-coll. & $(1,\theta^2, \theta)$ & $(1,\theta^2, \theta)$ & $(1,\theta^2, \theta)$ \\
     $\bn$-coll. & $(\theta^2,1, \theta)$ & $(\theta^2,1, \theta)$ & $(\theta^2,1, \theta)$ \\
     $n$-coll$_2$ & $(1,R^2,{R})$ & & & $(1,R^2, {R})$ \\
     $\bn$-coll$_2$ & $(R^2,1,{R})$ & & & $(R^2,1, {R})$ \\
     $n$-csoft & & & & $\theta/{R}(1,R^2, {R})$\\
     $\bn$-csoft &  & & & $\theta/{R}(R^2,1,{R})$ \\ 
     soft & $(\theta,\theta,\theta)$ & $(\theta,\theta,\theta)$ & $(\theta,\theta,\theta)$ & $(\theta,\theta,\theta)$ & $(\theta,\theta,\theta)$\\
     \hline \hline
   \end{tabular}
   \caption{The parametric scaling of the momenta $(p^-,p^+,\vecb{p})$ corresponding to the modes in SCET, for the various hierarchies between $\theta$ and ${R}$. For $\theta \ll {R}$ the modes differ between the Winner-Take-All and standard jet axis. All the momenta are normalized to $\sqrt{s}/2$.}
   \label{tab:modes}
   \end{table}

%===============================================================================
\subsection{$R \sim \theta \ll 1$}
\label{sec:fac1}
%===============================================================================

We now turn to the factorization analysis, starting with dijet production in $e^+e^-$ scattering at a center-of-mass energy $\sqrt{s}$, where $\theta \approx 2q_T/\sqrt{s} \sim {R} \ll1$. This is the simplest case since there are only two scales, $\sqrt{s}$ and $q_T$.     
The cross section differential 
in the momentum decorrelation $\vecb{q}$ and the jet energy fractions $z_i = 2E_{J,i}/\sqrt{s}$ factorizes as%
\footnote{In ref.~\cite{Gutierrez-Reyes:2018qez}, we denoted the $e^+e^-$ center-of-mass energy by $Q^2$, which we reserve for DIS in this paper. Furthermore, the argument of the jet function was missing the factor of $\tfrac12$ in front of $\sqrt{s}{\cal R}$.}
%%%
\begin{align}\label{eq:smallRfactorization}
   \frac{\df \si_{e^+e^-\rightarrow JJX}}{\df z_1\, \df z_2\, \df \vecb{q}} &= \sigma_{0}^{e^+e^-}(s)\, H_{e^+e^-}(s,\mu)
   \int \! \frac{\df \vecb{b}}{(2\pi)^2}\, 
   e^{-\img \textit{\textbf{b}}\cdot\textit{\textbf{{q}}}} 
   J_q(z_1, \vecb{b}, \tfrac{\sqrt{s}}{2}{\cal R},\mu, \zeta)\, J_{\bar q}(z_2, \vecb{b}, \tfrac{\sqrt{s}}{2}{\cal R},\mu, \zeta)
   \nn \\ & \quad \times
   \bigg[1 + \mathcal{O}\Big(\frac{{q}_T^2}{s}\Big)\bigg]
\,.\end{align}
%%%
The hard function $H_{e^+e^-}$ encodes the hard scattering process, in which a quark-anti-quark pair is produced. It contains virtual corrections, but no real radiation because that would result in $q_T \sim \sqrt{s}$. For convenience we have extracted the tree-level cross section $\sigma_{0}^{e^+e^-}$, which contains a sum over quark flavors. The jet functions describe the fraction $z_i$ of energy of the initial (anti)-quark that goes into the jet, as well as their transverse momentum through the impact parameter $\vecb{b}$ (its Fourier conjugate). They depend on the jet algorithm, as indicated by the argument $\tfrac{\sqrt{s}}{2}{\cal R}$, but this does not affect their anomalous dimension, as required by RG consistency. Soft radiation does not resolve the jet because its typical angle is order 1, whereas $R \ll 1$. Consequently, we do not have to consider clustering soft radiation in the jet algorithm, and we can simply include its effect as an overall recoil of the system, as indicated in \eq{smallRfactorization}. The soft function has been absorbed into the jet functions in the above expression, as we will discuss in \sec{resum}. There we will also show that the RG evolution between the hard scale $\mu_H \sim \sqrt{s}$ and jet scale $\mu_J \sim q_T$ in \eq{smallRfactorization} resums invariant mass logarithms of $\mu_H/\mu_J \sim \sqrt{s}/q_T$, and similarly that $\zeta$ is related to the resummation of invariant rapidity logarithms of $\sqrt{s}/q_T$~\cite{Scimemi:2017etj,Scimemi:2018xaf},  see also refs.~\cite{Collins:1984kg,Becher:2010tm,Collins:2011zzd,Chiu:2011qc,Chiu:2012ir}.

The corresponding factorization theorem for the cross section of semi-inclusive deep-inelastic scattering is given by
%%%
\begin{align}\label{eq:smallRfactorizationSIDIS}
   \frac{\df \si_{ep\rightarrow eJX}}{\df Q^2\, \df x\, \df z\, \df \vecb{q}} &= 
   \sum_q \sigma_{0,q}^{\rm{DIS}}(x,Q^2)\,H_{\rm DIS}(Q^2,\mu)
   \int \! \frac{\df \vecb{b}}{(2\pi)^2}\, e^{-\img \textit{\textbf{b}}\cdot\textit{\textbf{{q}}}} 
   F_{q}(x, \vecb{b}, \mu, \zeta)\, J_q\Big(z, \vecb{b}, \frac{Q{\cal R}}{2},\mu, \zeta\Big)
   \nn \\ & \quad \times
   \bigg[1 + \mathcal{O}\Big(\frac{{q}^2_T}{Q^2}\Big)\bigg]
\,,\end{align}
%%%
which is differential in the di-lepton invariant mass $Q^2$, Bjorken $x$, the energy fraction $z$ of the jet generated by the splitting of the quark, and the jet transverse momentum $q_T$. We work in the Breit frame, where $z=2E_J/Q$, and apply an $e^+e^-$ jet algorithm. 
The modification to the factorization theorem compared to \eq{smallRfactorization} is fairly modest: the hard function is replaced by the one for SIDIS, one of the jet functions is replaced by a TMD PDF, and the sum over quark flavors must be explicitly included because both $\sigma_{0,q}^{\rm{DIS}}$ and $F_q$ depend on it ($J_q$ does not, as long as we can treat quarks as massless).
The hard function is slightly different,
%%%
\begin{align}
H_{e^+e^-}(Q^2,\mu)&= |C_V(Q^2,\mu)|^2=1+2a_s C_F\left(-\mathbf{l}_{Q^2}^2-3\mathbf{l}_{Q^2}-8+\frac{7\pi^2}{6}\right)+\ord{a_s^2},
\nn \\
H_{\rm DIS}(Q^2,\mu)&= |C_V(-Q^2,\mu)|^2=1+2a_s C_F\left(-\mathbf{l}_{Q^2}^2-3\mathbf{l}_{Q^2}-8+\frac{\pi^2}{6}\right)+\ord{a_s^2}\,,
\end{align}
%%%
where $C_V$ is the Wilson coefficient for the hard matching, $\mathbf{l}_{Q^2}=\ln(\mu^2/Q^2)$ and $a_s=g^2/(4\pi)^2$. The NNLO and NNNLO expression can be found in ref.~\cite{Gehrmann:2010ue}, taking into account that $H_{e^+e^-}(Q^2,\mu)$ is the same as for the Drell-Yan process. The two loop  expressions are provided in eqs.~(\ref{eq:He+e-}, \ref{eq:HDIS}) of the appendix, to make the paper self-contained.

%===============================================================================
\subsection{${R} \ll \theta \ll 1$}
\label{sec:fac2}
%===============================================================================

We now consider the case where we have an additional hierarchy due to the small size of the jet radius, ${R} \ll \theta \ll 1$. While this regime will be of limited phenomenological interest to us, it allows us to make contact between our framework and TMD measurements with final state hadrons, corresponding to the ${R} \to 0$ limit. The modes are again listed in table~\ref{tab:modes}, and involve additional collinear modes whose scaling is set by ${R}$.

The factorization in this case is an extension of \eqs{smallRfactorization}{smallRfactorizationSIDIS}. The jet function contains two scales $\sqrt{s}{\cal R} \ll q_T$, which can be separated through a further collinear factorization,
%%%
\begin{align} \label{eq:smallR}
  J_i(z, \vecb{b}, E{\cal R}, \mu, \zeta)&=
  \sum_j\! \int\! \frac{\df z'}{z'}\, \big[(z')^2\mathbb{C}_{i \to j}(z', \vecb{b}, \mu, \zeta)\big]
   {\cal J}_j\Big(\frac{z}{z'},  \frac{2 z}{z'} E {\cR},\mu\Big)\,\big[1 \!+\! \mathcal{O}(b_T^2 E^2{\cal R}^2)\big]\,.
 \end{align}
%%%
Only collinear radiation at angular scales $\theta$, encoded in $\mathbb{C}_{i \to j}$, can affect $q_T$. However, subsequent splittings down to angles of order ${R}$ will change the parton $j$ with momentum fraction $z'$ into a jet with momentum fraction $z$. This is described by the semi-inclusive jet function ${\cal J}_j$, which has been calculated to $\ord{\al_s}$ in refs.~\cite{Kang:2016mcy,Dai:2016hzf} (our notation matches that of ref.~\cite{Kang:2016mcy}). The distinction between WTA vs.~standard jet axis is irrelevant, since $\theta \gg {R}$. The additional RG evolution between $\mu_J \sim q_T$ and $\mu_{\cal J} \sim E{\cal R}$ sums single logarithms of $\mu_J/\mu_{\cal J} \sim q_T/(E{\cal R}) \sim \theta/{R}$.

The $(z')^2$ in front of $\mathbb{C}_{i \to j}$ was chosen to ensure that these matching coefficients coincide with those for TMD fragmentation, given to $\ord{\al_s^2}$ in refs.~\cite{Echevarria:2015usa,Echevarria:2016scs}. That these same matching coefficients enter here is not surprising, since for ${R}\to 0$ the semi-inclusive jet function becomes the fragmentation function (summed over hadron species)~\cite{Kang:2017mda}. Thus in this limit we reproduce the known results for TMD fragmentation to hadrons. For convenience, we collect the relevant one-loop expressions for the matching coefficients  and semi-inclusive jet function in~\eqs{QsiJF}{QTMDmatching} of the appendix.

%===============================================================================
\subsection{$ \theta \ll {R}$ for the Winner-Take-All axis}
\label{sec:fac34_wta}
%===============================================================================
We now consider $\theta \ll {R}$ for the Winner-Take-All axis. For ${R} \sim 1$, the modes in table~\ref{tab:modes} are expected and factorization takes on a rather simple form. Even if soft radiation sees the jet boundary, it does not affect the position of the jet axis, due to the WTA recombination scheme. Specifically, the merging prescription in \eq{mergingPrescription} implies that soft radiation never affects the direction of the jet (it always ``loses'' against collinear radiation), while its contribution to the jet energy is power suppressed. The only effect of soft radiation, either inside or outside the jet, is thus therefore a total recoil of the two collinear sectors, which is therefore described by the standard TMD soft function. In particular, the observable is insensitive to the distinction between soft radiation inside and outside the jet. Since $\theta \ll R$, the collinear modes do not resolve the jet boundary, so $z=1$ and the $E {\cal R}$ dependence drops out,
%The factorization takes on a rather simple form, since soft radiation does not affect the position of the jet axis, due to the WTA recombination scheme. In particular, there is no distinction between soft radiation inside and outside the jet. The only effect of soft radiation is its total recoil, which is therefore described by the standard TMD soft function. Since $\theta \ll R$, the collinear modes do not resolve the jet boundary, so $z=1$ and the $E {\cal R}$ dependence drops out,
%%%
\begin{align} \label{eq:J_wta}
  J_i^{\rm WTA}(z, \vecb{b}, E {\cal R}, \mu, \zeta)
  &= \de(1-z)\, {\mathscr{J}}_i^{\rm WTA} (\vecb{b}, \mu, \zeta)
  \Big[1 + \mathcal{O}\Big(\frac{1}{b_T^2 E^2 {\cal R}^2}\Big)\Big]
\,.\end{align}
%%%
For completeness we also provide a definition of ${\mathscr{J}}_q^{\rm WTA}$, 
%%%
\begin{align} \
{\mathscr{J}}_q^{\rm WTA}(\vecb{b}) &=\frac{1}{2N_c}\text{Tr}\Big[\frac{\bnslash}{2} 
\langle 0|\Big(\frac{1}{\bn \sdt P}\,
e^{\img \textit{\textbf{b}}\cdot\textit{\textbf{P}}} \chi_n(0) \Big)
\sum_X |J_{\text{WTA}} \rangle \langle J_{\text{WTA}}|\bar\chi_n(0)|0\rangle\Big]
,\end{align}
%%%
and a similar formula can be written for the gluon case.

For $\theta \ll {R} \ll 1$, one would expect the same modes as are listed for the standard jet axis in table~\ref{tab:modes}. In this case the soft function does not resolve the jet boundary, because ${R} \ll 1$, but collinear-soft modes with scaling
%%%
\begin{align}
   (p^-, p^+, \vecb{p}) \sim \theta/R(1,R^2, R)\,, \quad \theta/R(R^2, 1,R)\,,
\end{align}
%%%
resolve the jet boundary and contribute to $q_T$. However, by the same reasoning as before, their only effect is a total recoil on the system, independent of whether emissions are inside or outside the jet. Consequently, these additional modes do not need to be considered, since they will simply be removed by the zero-bin subtraction~\cite{Manohar:2006nz}, due to their overlap with the soft mode. This leads to the interesting conclusion that, {\bf for the WTA axis, the cross section for $\theta \ll {R}$ is independent of ${R}$}.

%===============================================================================
\subsection{$\theta \ll R$ for the standard jet axis}
\label{sec:fac3_sja}
%===============================================================================

For completeness we also discuss $\theta \ll {R}$ for the standard jet axis. We do not present any numerical results for this case, and therefore limit our discussion to the dijet momentum decorrelation in $e^+e^-$ collisions. First we consider the case $\theta \ll {R} \sim 1$, for which the modes are given in table~\ref{tab:modes}. Energetic emissions outside the jet are not allowed because these would lead to $\theta \sim {R}$. Because the standard jet axis is along the total momentum of the jet, momentum conservation implies that $q_T$ is simply determined by the transverse momentum of soft radiation outside the jets. In particular, the angle of energetic emissions inside the jet is unrestricted. Since ${R}\sim 1$, these emissions are hard, explaining the absence of a collinear mode. Each of these hard emissions induces a soft Wilson line, implying the presence of non-global logarithms (NGLs)~\cite{Dasgupta:2001sh} of $\sqrt{s}{\cal R}/q_T$. The corresponding cross section can be described using the framework of refs.~\cite{Becher:2015hka,Becher:2016mmh} (see also refs.~\cite{Larkoski:2015zka,Caron-Huot:2015bja})
%%%
\begin{align}
  \frac{\df \si^{\rm SJA}_{e^+e^-\rightarrow JJX}}{ \df \vecb{q}} 
 &= 
    \sum_{m=2}^\infty {\rm Tr}_c[ \mathcal{H}_m(\{n_i\},\sqrt{s}, {\cal R}) 
    \otimes \mathcal{S}_m(\{n_i\},\vecb{q},{\cal R})] \bigg[1 + \mathcal{O}\Big(\frac{q_T^2}{Q^2}\Big)\bigg]
.\end{align}
%%%
We have eliminated the measurement of the momentum fractions of the jets, since  $z_i = 1$ in this limit.
$\mathcal{H}_m$ denotes the hard function with $m$ real emissions inside the jets, along the light-like directions $n_i$. The soft function $\mathcal{S}_m$ describes the transverse momentum $q_T$ of soft radiation outside the jets, produced by the Wilson lines along the directions $n_i$. The color indices describing the representation of the hard emissions/Wilson lines connects the hard and soft function, and ${\rm Tr}_c$ denotes the trace over these color indices. Finally, $\otimes$ denotes integrals over the light-like directions $n_i$.

Moving on to $\theta \ll {R} \ll 1$, we have collinear modes whose angular size is set by ${R}$, and additional collinear-soft modes with scaling
%%%
\begin{align}
   (p^-, p^+, \vecb{p}) \sim \theta/R(1,R^2, R)\,, \quad \theta/R(R^2, 1,R)\,,
\end{align}
%%%
which are fixed by the requirement that they resolve the jet boundary and contribute to $q_T$. Because ${R} \ll 1$, no hard real emissions are allowed, and the soft function does not resolve the jet. However, each collinear emission produces a collinear-soft Wilson line, in direct analogy to the soft Wilson lines generated by hard emissions for ${R} \sim 1$. Using again the framework of refs.~\cite{Becher:2015hka,Becher:2016mmh}, the corresponding cross section is given by 
%%%
\begin{align}
  \frac{\df \si^{\rm SJA}_{e^+e^-\rightarrow JJX}}{ \df \vecb{q}} 
 &= \sigma_{0}^{e^+e^-}(s)\, H_{e^+e^-}(s,\mu)  \int \! \frac{\df \vecb{b}}{(2\pi)^2}\, e^{-\img \textit{\textbf{b}}\cdot\textit{\textbf{{q}}}} \, S(\vecb b)
\nn \\ & \quad \times     
    \bigg[\sum_{m=2}^\infty {\rm Tr}_c[ \mathcal{J}_m(\{n_i\},\tfrac{\sqrt{s}}{2}{\cal R})
    \otimes \mathcal{U}_m(\{n_i\},\vecb{b},{\cal R})]\bigg]^2 
    \bigg[1 + \mathcal{O}\Big(\frac{q_T^2}{Q^2}\Big)\bigg]
.\end{align}
%%%
The hard and soft function are the same as for $\theta \sim R$. The jet function $\mathcal{J}_m$ describes $m$ collinear emissions inside a jet along light-like directions $n_i$, and the collinear-soft function $\mathcal{U}_m$ describes the resulting $q_T$ from collinear-soft emissions of these Wilson lines.

%%%%%%%%%%%%%%%%%%%%%%%%%%%%%%%%%%%%%%%%%%%%%%%%%%%%%%%%%%%%%%%%%%%%%%%%%%%%%%%%
\section{Quark jet function at one loop}
\label{sec:jet}
%%%%%%%%%%%%%%%%%%%%%%%%%%%%%%%%%%%%%%%%%%%%%%%%%%%%%%%%%%%%%%%%%%%%%%%%%%%%%%%%

In this section we present a detailed calculation of the one-loop quark jet function that enters the factorization formula in \eqs{smallRfactorization}{smallRfactorizationSIDIS}. We use dimensional regularization with $d = 4-2\ve$ to handle UV divergences, and the modified $\delta$-regulator for the rapidity divergences~\cite{Chiu:2009yx,Echevarria:2015usa},
The Feynman diagrams and measurement are discussed in \sec{jet_calc_intro}. We present a detailed calculation in momentum space in \sec{jet_pt} and in impact-parameter space in \sec{jet_bt}, thus providing a cross check of our results. The advantage of performing the calculation in momentum space is that this is the space in which the jet algorithm is defined. On the other hand, the renormalization and resummation are simpler in impact-parameter space. 

%===============================================================================
\subsection{Feynman diagrams and measurement}
\label{sec:jet_calc_intro}
%===============================================================================

%%%
\begin{figure*}[tb]
  \centering
  \includegraphics[width=0.8\textwidth]{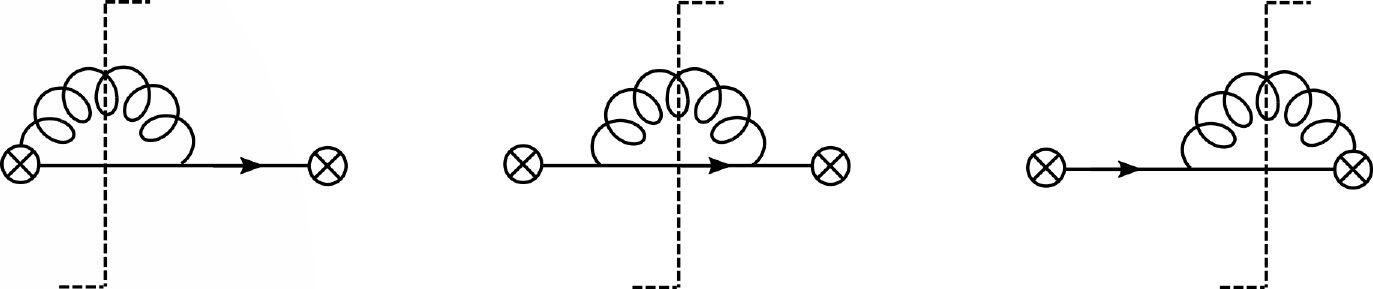}
  \caption{Cut diagrams that contribute to the one-loop quark jet function in SCET. Here $\otimes$ represents the collinear (anti-)quark field $\chi_n$ ($\bar \chi_n$), which contains a collinear Wilson line that can emit gluons. A sum over cuts is understood, where cuts through loops describe real emissions, while only cutting the quark line corresponds to virtual corrections. The latter vanish for our choice of regulators.}
\label{fig:diagrams}
\end{figure*}
%%%
The one-loop diagrams that contribute to the quark function are given in \fig{diagrams}. This leads to the following expression for the bare jet function up to one loop,
%%%
\begin{align} \label{eq:quarkFunctionStart}
J^{\rm alg}_{q}(z,\vecb{q}, E\mathcal{R}) &= \sum_n a_s^nJ_q^{[n]} (z,\vecb{q}, E\mathcal{R})
 \\ &
=
 \frac{1}{\pi} \delta(q_T^2)\delta(1-z) 
 \nn \\ & \quad
 + g^2\sum_{\text{cases}}C_F\Big(\frac{\mu^2e^{\ga_E}}{4\pi}\Big)^\ve\! \int_0^\infty \!\!\!\frac{\df \ell^+}{2\pi\ell^+} \!\int\! \frac{\df^d k}{(2\pi)^d} \bigg[ 2\frac{2E\!-\!k^-}{k^- \!-\! \img\de^-}+(1\!-\!\ve)\Big(1\!-\!\frac{k^+}{\ell^+}\Big) + \text{h.c.} \bigg]
  \nn \\ & \quad
  \times (2\pi)\de^+(k^2)\,(2\pi)\de^+\big[(\ell-k)^2\big]\, \Thh_{\case}\,\frac{1}{\pi}\de(q_T^2\!-\!q^2_{T\text{case}})\,\de\Bigl(z\!-\!\frac{E_{J\,\text{case}}}{E}\Bigr) + \ord{a_s^2}. \nn
\end{align}
%%%
Here $E$ is the energy of the quark field initiating the jet, and its small light-cone component $\ell^+$ (and thus  virtuality) is integrated over. The phase space of the outgoing gluon, with momentum $k^\mu$, and quark, with momentum $\ell^\mu - k^\mu$ is integrated over, subject to the $q_T$ and $z$ measurement. The $\de^+(k^2) \equiv \de(k^2)\thh(k_0)$ and $\de^+[(k-\ell)^2]$ denote the corresponding on-shell conditions. The coupling has been replaced by the renormalized one in the $\overline{\text{MS}}$ scheme, leading to the prefactor $(\dots)^\ve$. 

There are three different cases we need to consider:
\begin{itemize}
 \item[(a)] both partons are inside the jet, \\[-4.5ex]
 \item[(b)] the gluon is outside the jet, \\[-4.5ex]
 \item[(c)] the quark is outside the jet.
\end{itemize}
These cases are identified by $\Thh_{\case}$, and the transverse momentum $q^2_{T\text{case}}$ and jet energy $E_{J,\text{case}}$ depend on the case and jet algorithm, and are given in table~\ref{tab:caseConstraints} in terms of the energy fraction of the quark
%%%
\begin{align}
  x \equiv 1-\frac{k^-}{2E}
\end{align}
%%%
and of the jet size $\mathcal{R}$, defined in eq.~(\ref{eq:jetsize}).

At one loop, there are only two partons, so every distance measure gives the same clustering condition (as we will see in \sec{jet_two}, this is no longer true at two loops). There are differences between the standard and WTA recombination scheme that directly follow from the different rules in \eq{mergingPrescription}. This distinction is only relevant when both partons are inside the jet, in which case the standard jet axis is along their total momentum while the WTA axis is along the most energetic one.
%%%
\begin{table}
\centering
  \begin{tabular}{l | c | c | c | c }
  case & algorithm & $\Thh_\case$ & $E_{J\,\case}$ & $q^2_{T\rm case}$\\
			\hline
  \multirow{2}{*}{(a) both in} & SJA & \multirow{2}{*}{$\theta\big(x(1-x)E\cR-k_T \big)$} & \multirow{2}{*}{$E$} & 0 \\ 
  & WTA & & & $\displaystyle \frac{k_T^2}{\max^2(x,1-x)}\phantom{\Bigg|}$ \\
  (b) gluon out	 & SJA/WTA & $\theta\big(k_T-x(1-x)E\cR \big)$	& $Ex$ & $\displaystyle \frac{k_T^2}{x^2}\phantom{\bigg|}$ \\
  (c) quark out & SJA/WTA & $\theta\big(k_T-x(1-x)E\cR \big)$	& $E(1-x)$ & $\displaystyle \frac{k_T^2}{(1-x)^2}$\\
  \end{tabular}
  \caption{The $\Thh_\case$ that encodes the various regions of phase space, and the corresponding jet energy $E_{J\,\case}$ and transverse momentum $q^2_{T\rm case}$. At this order the only difference between jet algorithms is the recombination scheme, i.e.~standard jet axis vs.~Winner-Take-All.}
  \label{tab:caseConstraints}
\end{table}
%%%

Switching from $k^-$ to the quark energy fraction $x$, using the on-shell conditions, and exploiting azimuthal symmetry, we rewrite the one-loop term of \eq{quarkFunctionStart} as 
%%%
\begin{align} \label{eq:quarkFunctionGeneral}
J^{\rm alg\,\one}_{q}(z,\vecb{q}, E\cR) &=  \sum_{\rm cases} \frac{4C_F}{\pi}\frac{\big(\mu^2e^{\ga_E}\big)^\ve}{\Ga(1-\ve)} \int_0^1 \df x  \int_0^\infty\!\! \frac{\df k_T}{k_T^{1+2\ve}} \,
 \Thh_{\case}\,\de(q_T^2-q^2_{T\text{case}})\,\de\Bigl(z-\frac{E_{J\,\text{case}}}{E}\Bigr)
  \nonumber \\ & \quad \times \bigg[\frac{(1+x^2)(1-x)}{(1-x)^2+(\de^-_E)^2}-(1-x)\ve\bigg]
 \,.\end{align}
%%%
Here we replaced the $\de^-$ regulator by its dimensionless counterpart 
%%%
\begin{align} \label{eq:dimensionlessDelta}
  \de^\pm_E \equiv \frac{\de^\pm}{2E}.
\end{align}
%%%
After similar manipulations, the corresponding one-loop gluon jet function is
%%%
\begin{align} \label{eq:gluonJetFunctionGeneral}
  J^{\rm alg\, \one}_{g}(z,\vecb{q}, E\cR) &= \sum_{\rm cases} \frac{4}{\pi}\frac{\big(\mu^2e^{\ga_E}\big)^\ve}{\Ga(1-\ve)} \int_0^1 \df x  \int_0^\infty\!\! \frac{\df k_T}{k_T^{1+2\ve}} \,
   \Thh_{\case}\,\de(q_T^2-q^2_{T\text{case}})\,\de\Bigl(z-\frac{E_{J\,\text{case}}}{E}\Bigr)\nonumber \\
  &\times \bigg\{ C_A(1-x)\Big[x+\frac{1}{x}+\frac{x}{(1-x)^2+(\de^-_E)^2}\Big] +n_fT_F\Big[1-\frac{2x(1-x)}{1-\ve}\Big] \bigg\}\, .
\end{align}
%%%
From this expression one can obtain the one-loop result for the gluon jet function presented in ref.~\cite{Gutierrez-Reyes:2018qez}, following step by step the calculation of the quark function detailed below.

%===============================================================================
\subsection{One loop results in momentum space}
\label{sec:jet_pt}
%===============================================================================

In order to perform the calculation in transverse momentum space we directly solve the two integrals in \eq{quarkFunctionGeneral}, inserting the measurements for the various cases in table~\ref{tab:caseConstraints}. We start with the case of both partons inside the jet. 

In the case of the standard jet axis, the dependence on the transverse momentum is trivial and the calculation reduces to the one performed in ref.~\cite{Kang:2016mcy} for the semi-inclusive quark jet function. After integration over the transverse momentum,
%%%
\begin{align}
J^{\rm SJA\,\one}_{q (a)} = -\frac{2C_F}{\pi} \Big(\frac{\mu^2}{E^2\cR^2}\Big)^\ve\! \frac{e^{\ve\ga_E}}{\ve\Ga(1-\ve)}\,\de(1-z)\,\de(q_T^2) \!\int_0^1\df x\, x^{-2\ve} (1-x)^{1-2\ve}\Big[\frac{1+x^2}{(1-x)^2}-\ve\Big].
\end{align}
%%%
Here we set the rapidity regulator $\de^-_E$ to zero because the endpoint $x=1$ is already regulated by dimensional regularization. The remaining integral over the energy fraction is a combination of Euler Beta functions, whose expansion up to $\ord{\ve^0}$ yields
%%%
\begin{align}\label{eq:EpairSJA}
J^{\rm SJA\,\one}_{q (a)} = \frac{2C_F}{\pi} \,\de(1-z)\,\de(q_T^2) \bigg[
	\frac{1}{\ve^2} + \frac{1}{\ve}\Big(L_R +\frac{3}{2}\Big) + \frac{1}{2}L_R^2 +\frac{3}{2}L_R +
	\frac{13}{2}-\frac{3\pi^2}{4}\bigg]\, ,
\end{align}
%%%
where 
%%%
\begin{align}
  L_R = \ln \Big(\frac{\mu^2}{E^2\cR^2}\Big)\,.
\end{align}
%%%

For the WTA axis, the transverse momentum dependence becomes nontrivial. The condition $\max(x,1-x)$ reduces to $x>\tfrac{1}{2}$ if we symmetrize the integrand,
%%%
\begin{align}\label{eq:sym}
J^{\rm WTA\,\one}_{q (a)} = \frac{2C_F}{\pi} \frac{e^{\ve\ga_E}}{\Ga(1-\ve)}\frac{\mu^{2\ve}}{(q_T^2)^{1+\ve}}\,\de(1-z) &\int_{\frac{1}{2}}^1\df x\, x^{-2\ve} \thh\big((1-x)E\cR-q_T\big)
\nn \\
	 &\times \bigg[\Big(-3+\frac{2}{x}-\ve\Big)+2\frac{1-x}{(1-x)^2+(\de^-_E)^2}\bigg]\, .
\end{align}
%%%
Performing the remaining integral requires to treat the integrand as a two-dimensional distribution, see \eq{two-dim_distribution}, and yields the result
%%%
\begin{align} \label{eq:EpairWTA}
J^{\rm WTA\,\one}_{q (a)} &= \frac{2C_F}{\pi} \de(1-z)\bigg\{\de(q_T^2) \bigg[
  \frac{1}{\ve^2}+\frac{1}{\ve} \Big(L_R+\frac{3}{2}\Big)+\frac{1}{2}L_R^2+\frac{3}{2}L_R+\frac{7}{2}-2\ln^2 2-\frac{5\pi^2}{12}\bigg]
  \nn\\ & \quad
  -\LoneQTR
  +\Big(2\ln 2-\frac{3}{2}\Big)\LzeroQTR
  \nn \\ & \quad
  +\thh\Big(\frac{E\cR}{2}-q_T\Big)\frac{1}{q_T^2}\bigg[3\frac{q_T}{E\cR}+2\ln\Big(1-\frac{q_T}{E\cR}\Big) \bigg]\bigg\}.
\end{align}
%%%

Finally we consider the cases where only one particle is inside the jet, that are independent of the jet algorithm. We use $x\rightarrow 1-x$ to combine the case where the gluon is outside the jet with the case where the quark is outside. Both the integrals over transverse momentum and energy fraction are fixed by the $\de$ functions enforcing the measurement, resulting in
%%%
\begin{align} \label{eq:EsingleUnexpanded}
J^{\one}_{q (b)+(c)} = \frac{2C_F}{\pi}\,\frac{\mu^{2\ve}}{(q_T^2)^{1+\ve}}\frac{e^{\ve\ga_E}}{\Ga(1-\ve)}\theta\Big(z-1+\frac{q_T}{E\cR}\Big)\Big[\Big(-3+\frac{2}{z}-\ve\Big)+\frac{2(1-z)}{(1-z)^2+(\de_E^-)^2}\Big]z^{-2\ve}.
\end{align}
%%%
Expanding the result in $\ve$ and $\de^-_E$ requires again some algebra with distributions, that is performed explicitly in \app{plus}. We obtain
%%%
\begin{align}\label{eq:Esingle}
J^{\one}_{q (b)+(c)} &= \frac{2C_F}{\pi} \bigg\{\de(q_T^2)\de(1-z) \Big[
  -\frac{1}{\ve^2} +\frac{1}{\ve} \Big(2\ln\de^-_E - L_R\Big) -\frac{1}{2}L_R^2+\frac{\pi^2}{12}\Big]
  \\ & \quad
  +\Big(-3+\frac{2}{z}+2\cL_0(1-z)\Big)\Big[\LzeroQT-\LzeroQTZ + L_R\de(q_T^2)\Big]
  \nn \\  &\quad
  -2\ln\de^-_E\,\LzeroQT\de(1-z)-2\de(q_T^2)\Big[\Big(-3+\frac{2}{z}\Big)\ln(1-z)+2\cL_1(1-z)\Big]\bigg\}.
\nn\end{align}
%%%
We now combine the expressions in eqs.~(\ref{eq:EpairSJA}) and~(\ref{eq:EpairWTA}) with~(\ref{eq:Esingle}), to obtain the bare quark jet function at one loop
%%%
\begin{align} \label{eq:bareEjetSJA}
J^{\rm{axis}\,\one}_{q} &=\nn \frac{2C_F}{\pi}\bigg\{\de(1-z)
  \Big[\de(q_T^2)\Big(\frac{2}{\ve}\ln\de^-_E+\frac{3}{2\ve}+\frac{3}{2}L_R\Big) -2\ln\de^-_E\,\LzeroQT +\Delta_q^{\rm axis}(q_T^2)\Big] 
  \\ & \quad +
  \big(p_{qq}(z)+p_{gq}(z)\big)\Big[\de(q_T^2)L_R+\LzeroQT-\LzeroQTZ\Big]
  \nn \\ & \quad
 -2\Big[\Big(-3+\frac{2}{z}\Big)\ln(1-z)+2\cL_1(1-z)\Big]\de(q_T^2)\bigg\}\, .
\end{align}
%%%
The dependence on the algorithm occurs via the functions $\Delta_q^{\rm axis}$, that explicitly read
%%%
\begin{align} \label{eq:axisDependentDeltas}
  \Delta_q^\SJA (q_T^2)& = \de(q_T^2) \Big(\frac{13}{2}-\frac{2\pi^2}{3}\Big)\, , \\
  \Delta_q^\WTA (q_T^2) & =\de(q_T^2) \Big(\frac{7}{2}-2\ln^2 2 - \frac{\pi^2}{3}\Big) 
  + \thetaQT\frac{1}{q_T^2}\bigg[\frac{3q_T}{E\cR}+2\ln\Big(1-\frac{q_T}{E\cR}\Big)\bigg]	
  \nn \\ & \qquad
 + \Big(2\ln 2-\frac{3}{2}\Big)\LzeroQTR -\LoneQTR\, .
\end{align}
%%%
The expression for the WTA axis is more involved because it introduces the threshold $z>\frac{1}{2}$. We notice that 
%%%
\begin{align}
  \Delta_q^\WTA(q_T^2) = \Delta_q^\SJA (q_T ^2)\,\bigg[1 + \cO\Big(\frac{E^2{\cal R}^2}{q_T^2}\Big)\bigg]\, .
\end{align}
%%%
This implies that the dependence on the jet algorithm vanishes in the regime $R \ll \theta$, as predicted from the factorization formula in \eq{smallR} (the semi-inclusive jet function $\mathcal{J}$ that enters there is independent of the jet axis).

%===============================================================================
\subsection{One-loop results in impact-parameter space}
\label{sec:jet_bt}
%===============================================================================

The calculation of the quark jet function at one loop can also directly be performed in impact-parameter space. This calculation provides a check of the results in the previous section. We perform the same two integrals of \eq{quarkFunctionGeneral} with the cases shown in the table~\ref{tab:caseConstraints} as in the momentum-space calculation, but first carry out the Fourier transform of the jet function
%%%
\begin{align}
\label{eq:Ftrans}
J_{q}^{\text{alg[1]}}(z,\vecb b,E{\cal R})=\int\! \df \vecb q\,e^{\img \textit{\textbf{b}}\cdot\textit{\textbf{{q}}}}\, J_{q}^{\text{alg[1]}}(z,\vecb q,E{\cal R}).
\end{align}
%%%
The case with both partons inside the jet is the only one that depends on the choice of axis. The result for SJA has a trivial dependence on the transverse momentum and can be written as
%%%
\begin{align}\label{eq:EpairSJAb}
J^{\rm SJA\,\one}_{q (a)} = 2C_F\,\de(1-z)\bigg[
	\frac{1}{\ve^2} + \frac{1}{\ve}\Big(L_R +\frac{3}{2}\Big) + \frac{1}{2}L_R^2  +\frac{3}{2}L_R +
	\frac{13}{2}-\frac{3\pi^2}{4}\bigg].
\end{align}
%%%
Note that for this calculation the IR divergences are regulated by $\ve$ and we can safely neglect the $\delta^-_E$ regulator.

The WTA axis choice introduces a non trivial dependence on the transverse momentum of the jet function. Symmetrizing the integral over $x$, as in eq.~\eqref{eq:sym}, we rewrite the jet function as 
%%%
\begin{align}\label{eq:EpairWTAb}
J^{\rm WTA\,\one}_{q (a)} =8\pi C_F&\Bigl(\frac{\mu^2 e^{\ga_E}}{4\pi}\Bigr)^\ve \delta(1-z)\int_{1/2}^1\! \df x \Big[\frac{1+x^2}{1-x}-\ve(1-x)+\frac{1+(1-x)^2}{x}-\ve x\Big]\nn\\
&\int\! \frac{\df^{d-2}\vecb k}{(2\pi)^{d-2}}\,\frac{1}{k_T^2}\,\theta\big(x(1-x)E\mathcal{R}-k_T\big)e^{\img \textit{\textbf{b}}\cdot\textit{\textbf{{k}}}/x}\,.
\end{align}
%%%
Integration over the transverse momentum allows us to rewrite \eq{EpairWTAb} as 
%%%
\begin{align}\label{eq:EpairWTAb2}
J^{\rm WTA\,\one}_{q (a)} &= 2C_F\frac{\big(\mu^2 e^{\ga_E}\big)^\ve}{\Gamma^2(1-\ve)} \delta(1-z)(E\mathcal{R})^{-2\ve}\Ga(-\ve)\int_{1/2}^1\! \df x\, x^{-2\ve}(1-x)^{-2\ve} 
\nn\\ &\qquad \times
\Big[\frac{1+x^2}{1-x}-\ve(1-x)+\frac{1+(1-x)^2}{x}-\ve x\Big]
\nn\\ & \quad
 -2C_F\delta(1-z)B_{ER}^2 \int_{1/2}^1 \df x \Big[(1+x^2)(1-x)+\frac{(1+(1-x)^2)}{x}(1-x)^2\Big]
 \nn\\ & \qquad \times
{_2}F_3\Big(1,1;2,2,2;-B_{ER}^2(1-x)^2\Big) + \ord{\ve}.
\end{align}
%%%
The jet function depends on the transverse position in terms of the dimensionless combination 
%%%
\begin{align}
B_{ER}=\frac12 b_T E\mathcal{R}
\,.\end{align}
%%% 
The remaining step is the integration over $x$. The integral in the first term (first two lines) is straightforward to perform analytically. On the other hand, the second integral has a part for which we were unable to obtain a closed analytical expression. The result of this second integral is given by the function $\mathcal{G}(B_{ER})$, whose explicit expression is given in \eq{cG}. This leads to
%%%
\begin{align}\label{eq:EpairWTAb3}
J^{\rm WTA\,\one}_{q (a)} &=2C_F\,\de(1-z)
\bigg[
	\frac{1}{\ve^2} + \frac{1}{\ve}\Big(L_R +\frac{3}{2}\Big) + \frac{1}{2}L_R^2  +\frac{3}{2}L_R +
	\frac{13}{2}-\frac{3\pi^2}{4} + {\cal G}(B_{ER}) \bigg]
\end{align}
%%%
Note that the only difference between the SJA and WTA results is $\cal G$. When $B_{ER}\ll 1$ the function $\mathcal{G}$ is zero, as required by the axis independence in this limit. 

Next we consider the case when only one parton is inside the jet. By using $x\to 1-x$, we can combine case (b) and (c). As we now have an explicit dependence on the momentum fraction of the jet,  the rapidity regulator $\delta^-_E$ needs to be kept. We find, 
%%%
\begin{align}\label{eq:EsingWTAb}
J^{\one}_{q (b)+(c)} &=2C_F\biggl\{\bigl(p_{qq}(z)+p_{gq}(z)\bigr)\biggl[L_R-L_\mu-2\ln(1-z)
\nn\\ & \qquad
+B_{ER}^2(1-z)^2{_2}F_3\Bigl(1,1;2,2,2;-B_{ER}^2(1-z)^2\Bigr)\biggr]
\nn\\ & \quad
+\delta(1-z)\Bigl[\frac{2}{\ve}\ln\de_E^-+2L_\mu\ln\de^-_E\Bigr]+\de(1-z)\Bigl[\frac{3}{2\ve} +\frac{3}{2}L_R+ \frac{13}{2}-\frac{2\pi^2}{3}\Bigr]
\nn\\ & \quad
-\de(1-z)\Bigl[
	\frac{1}{\ve^2} + \frac{1}{\ve}\Big(L_R +\frac{3}{2}\Big) + \frac{1}{2}L_R^2  +\frac{3}{2}L_R +
	\frac{13}{2}-\frac{3\pi^2}{4}\Bigr]\biggr\}.
\end{align}
%%%
The terms with a divergent behavior in the limit $z\to 1$ should be understood as regulated under the +-prescription. For clarity we have split the $\de(1-z)$ contribution into three pieces: the first term will be eliminated after the renormalization of rapidity divergences, and the third term is exactly cancelled by the corresponding part of the case with both particles inside the jet, removing IR divergences presented here as double poles in $\ve$.

The final result for the quark jet function for both choices of axis is obtained summing eq.~\eqref{eq:EpairSJAb} (SJA) or \eqref{eq:EpairWTAb3} (WTA) with \eqref{eq:EsingWTAb},
%%%
\begin{align}\label{eq:bspaceFINAL}
J^{\rm axis\,\one}_{q} &=2C_F\bigg\{\bigl(p_{qq}(z)+p_{gq}(z)\bigr)\biggl[L_R-L_\mu-2\ln(1-z)
\nn\\ & \quad 
+B_{ER}^2(1-z)^2{_2}F_3\Bigl(1,1;2,2,2;-B_{ER}^2(1-z)^2\Bigr)\biggr]
\nn\\ & \quad
+\delta(1-z)\Bigl(\frac{2}{\ve}\ln\de^-_E+2L_\mu\ln\de^-_E\Bigr)+\de(1-z)\Bigl(\frac{3}{2}L_R+\frac{3}{2\ve}+\tilde \Delta_q^{\rm axis}(B_{ER})\Bigr)\bigg\},
\end{align}
%%%
where 
%%%
\begin{align}\label{eq:axisdep}
\tilde \Delta_q^{\rm SJA}(B_{ER})=\frac{13}{2}-\frac{2\pi^2}{3},
\, \qquad
\tilde \Delta_q^{\rm WTA}(B_{ER})=\frac{13}{2}-\frac{2\pi^2}{3}+\mathcal{G}(B_{ER}).
\end{align}
%%%
We have checked that these expressions agree with those obtained in \sec{jet_pt}, which is partially numerical for the WTA axis. For the numerical implementation, we use the above expressions when a closed analytic expression is available, while we find it more convenient to estimate the sum $\mathcal{S}$, defined in \eq{Sapprox}, by numerically Fourier transforming its momentum-space counterpart.

%%%%%%%%%%%%%%%%%%%%%%%%%%%%%%%%%%%%%%%%%%%%%%%%%%%%%%%%%%%%%%%%%%%%%%%%%%%%%%%%
\section{Renormalization and resummation}
\label{sec:resum}
%%%%%%%%%%%%%%%%%%%%%%%%%%%%%%%%%%%%%%%%%%%%%%%%%%%%%%%%%%%%%%%%%%%%%%%%%%%%%%%%

%===============================================================================
\subsection{Rapidity renormalization}
\label{sec:rap_ren}
%===============================================================================

The jet function in \eq{Jq_def} has the same renormalization as in the case of TMDs. Here we summarize the main points of rapidity renormalization, referring to e.g.~ref.~\cite{Echevarria:2016scs} for further details. The rapidity renormalization factor $R_q$ in \eq{ZR} can be extracted from the soft function 
%%%
\begin{align}
R_q(\vecb{b},\zeta,\mu)=\frac{\sqrt{S_q(\vecb{b},\zeta,\mu)}}{\textbf{Zb}_q},
\end{align}
%%%
including the zero-bin $\textbf{Zb}_q$, that accounts for the overlap with collinear modes. 

The soft function for SIDIS is given by the following vacuum matrix element of soft Wilson lines
%%%
\begin{align} \label{eq:SF_def}
S_q(\vecb b)
= \frac{1}{N_c}\,{\rm Tr}_c \sandwich{0}{\bar T\bigl[\tilde S^{\dagger}_\bn S_n \bigr](0^+,0^-,\vecb b)T\bigl[S_n^{\dagger} \tilde S_\bn \bigr] 
(0)}{0},
\end{align}
%%%
where the coordinates in brackets indicate the position of both Wilson lines, and $T$ ($\bar T$) denotes (anti-)time ordering.
The Wilson lines are defined as usual 
%%%
\begin{align}
 \label{eq:SF_def2}
S_n (x) &= P \exp \biggl[\img g \int_{-\infty}^0\! \df \sigma\, n \sdt A (x+\sigma n)\biggr]
,\\ \nn
\tilde S_\bn (x) &= P\exp\biggl[-\img g\int_{0}^{\infty}\! \df \sigma\, \bn \sdt A(x+ \sigma\bn) \biggr]
.\end{align}
%%%
In \eq{SF_def} we did not include the transverse gauge links which are necessary to preserve the  gauge invariance in singular gauges~\cite{Belitsky:2002sm,Idilbi:2010im,GarciaEchevarria:2011md}, because we do not use them in our computations. For  $e^+e^-$ all Wilson lines are future pointing, which corresponds to $\tilde S_\bn \to S_\bn$ in \eq{SF_def}, but this fact has no practical consequences as in the case of TMDs \cite{Collins:2011zzd,Echevarria:2012js,Echevarria:2014rua,Vladimirov:2017ksc}.
The overlap of collinear and soft modes depends in general on the rapidity regulator used in the perturbative calculation and for the {\it{modified}} $\delta$-regulator used in the present work one finds that in fact $S_q(\vecb{b},\zeta,\mu)=\textbf{Zb}_q$, so that the  rapidity renormalization factor has the simple form $R_q(\vecb{b},\zeta,\mu)=1/\sqrt{S_q(\vecb{b},\zeta,\mu)}$. 

The parameter $\zeta$ in $R_q$ is a scale that comes from splitting  the soft function in two factors, 
%%%
\begin{align}
S\biggl(\vecb{b};\ln\Bigl(\frac{\mu^2}{\delta^+\delta^-}\Bigr)\biggr)=S^{1/2}\biggl(\vecb{b};\ln\Bigl(\frac{\mu^2}{(\delta_E^+)^2\zeta_+}\Bigr)\biggr)
S^{1/2}\biggl(\vecb{b};\ln\Bigl(\frac{\mu^2}{(\delta_E^-)^2\zeta_-}\Bigr)\biggr)
\end{align}
%%%
each one of which is absorbed in one of the jet functions or TMDs. Specifically, the perturbative calculation of the soft function reveals that it depends linearly on $\ln(\mu^2/(\delta^+\delta^-))$, where the $\de^\pm$  are the rapidity regulators for each of the collinear modes in the factorization theorem. 
To separate them, $\zeta_\pm$ are introduced with $\zeta_+\zeta_-=(2 E)^4$, and $2E$ is the hard scale of the process under consideration. In the calculation of a jet function along the direction $n$ one can effectively replace 
$\delta_E^-=\delta_E^+\zeta$, so that the subscripts $\pm$ for the variable $\zeta$ can be omitted.
While the rapidity renormalization factor is simply multiplicative in $\vecb b$-space, the jet function can also be calculated in momentum space, as we have shown in the previous section.

%===============================================================================
\subsection{One-loop renormalization of the jet function and small and large $R$ limits}
%===============================================================================

Our bare jet function in \eqs{bareEjetSJA}{bspaceFINAL} is still affected by divergences. As discussed in \eq{ZR} and \sec{rap_ren}, its renormalization is particularly easy to implement in impact-parameter space, where it is purely multiplicative and takes the same form as for hadron TMDs. The explicit one-loop UV and rapidity renormalization factors are 
%%%
\begin{align}
  Z_q^\one(\zeta,\mu) &= -\frac{2}{\ve}C_F\Big(\frac{1}{\ve}+\textbf{l}_\zeta+\frac{3}{2}\Big)\, , \\
  R_q^\one(\zeta,\mu) &= 2C_F\bigg[\frac{1}{\ve^2}-\Big(\frac{1}{\ve}+L_\mu\Big)\ln\Big(\frac{(\delta_E^-)^2\zeta}{\mu^2}\Big)
  -\frac{1}{2}L_\mu^2-\frac{\pi^2}{12}\bigg]\, ,
\end{align}
%%%
leading to the renormalized expression 
%%%
\begin{align}\label{eq:bspaceRenorm}
J^{\rm axis\,\one}_{q} (z,\vecb b,E{\cR},\mu,\zeta)&=2C_F\bigg\{\bigl(p_{qq}(z)+p_{gq}(z)\bigr)\biggl[L_R-L_\mu-2\ln(1-z)
\nn\\ & \quad 
+B_{ER}^2(1-z)^2{_2}F_3\Bigl(1,1;2,2,2;-B_{ER}^2(1-z)^2\Bigr)\biggr]
\nn\\ & \quad
+\delta(1-z)\Bigl(L_\mu\textbf{l}_\zeta-\frac{1}{2}L_\mu^2+\frac{3}{2}L_R+\tilde \Delta_q^{\rm axis}(B_{ER})-\frac{\pi^2}{12}\Bigr)\bigg\}\, .
\end{align}
%%%
The corresponding momentum-space result is presented in \eq{JetFunction_q}.

From \eq{bspaceRenorm} (or equivalently from \eq{JetFunction_q}) one can take the limits $\mathcal{R}\to0$ and $\mathcal{R}\to \infty$, to approach the factorization regimes described respectively in \sec{fac2} and in \secs{fac34_wta}{fac3_sja}. In the small-$R$ limit the two axes give the same result, and we explicitly checked that the jet function factorizes further as in \eq{smallR}. The perturbative ingredients in which the jet function factorizes are listed in \app{fixed_order}. The large-$R$ limit is particularly interesting for the WTA axis, where the jet function simplifies as in \eq{J_wta}. We verified that the dependence on the jet radius drops out in this limit, obtaining
%%%
\begin{align}
  \mathscr{J}^{\rm{WTA}\one}(\vecb{b},\mu,\zeta) = 2C_F\Big(\frac{7}{2}-\frac{5\pi^2}{12}-3\ln 2 -\frac{1}{2}L_\mu^2+
   L_\mu\textbf{l}_\zeta+\frac{3}{2}L_\mu  \Big)\, .
\end{align}
%%%

%===============================================================================
\subsection{Resummation and $\zeta$-prescription}
\label{sec:jet_ren}
%===============================================================================

The renormalization group equations (RGEs) of the TMD jet function are the same as for the standard hadronic TMD,
%%%^
\begin{align}
\label{RGE}
\mu \frac{\df}{\df\mu}J_q(\vecb b;\mu,\zeta)&= \gamma_q(\mu,\zeta)J_q(\vecb b;\mu,\zeta)
\nn \\
\zeta \frac{\df}{\df\zeta}J_q(\vecb b;\mu,\zeta)&=-\mathcal{D}_q(\mu;\vecb{b})J_q(\vecb b;\mu,\zeta)
\end{align}
%%%
where $\mathcal{D}_q$ and $\ga_q$ are the rapidity and UV anomalous dimension, respectively. We only consider the quark jet function, because the gluon does not enter in our phenomenological results.
As in the hadronic TMD case we have
%%%
\begin{align}
\mathcal{D}_q=-\frac{\df \ln R_q}{\df \ln\zeta}\Big|_{f.p.}=-\frac{1}{2}\frac{\df \ln R_q}{\df \ln \delta^+}\Big|_{f.p.},
\end{align}
%%%
where $|_{f.p.}$ denotes the finite parts.

Since the order of derivatives can be interchanged, one obtains~\cite{Chiu:2011qc,Echevarria:2015byo},
%%%
\begin{align}
\mu \frac{\df}{\df\mu}\(-\mathcal{D}_q(\mu^2,\vecb{b})\)=\zeta\frac{\df}{\df\zeta}\gamma_q(\mu,\zeta) = - \Gamma^{\rm cusp}_q.
\label{eq:cusp1}
\end{align}
%%%
where $\Gamma^{\rm cusp}_q$ is the quark cusp anomalous dimension.
Consequently,
%%%
\begin{align}
\gamma_q=\Gamma^{\rm cusp}_q\mathbf{l}_\zeta-\gamma_{V,q},
\end{align}
%%%
where
%%%
\begin{align}\label{def_logarithms}
\mathbf{l}_\zeta\equiv \ln\(\frac{\mu^2}{\zeta}\)\,,
\end{align}
%%%
and $\gamma_V$ is the finite part of the renormalization of the vector form factor. 
Both $\gamma_V$ and $\mathcal{D}$ 
are known up to $\ord{a_s^3}$~\cite{Moch:2004pa,Moch:2005tm,Baikov:2009bg,Vladimirov:2016dll,Li:2016ctv}, and a numerical computation of the fourth-order cusp anomalous dimension was recently presented in ref.~\cite{Vogt:2018miu}. All these anomalous dimensions are collected in \app{anom_dim}.

The high-energy scale value for $\mu$ is always set at the hard scale, i.e.~$\sqrt{s}$ for $e^+e^-$ and $Q$ for SIDIS.
As for the TMD case, the  evolution of the  jet function in the plane $(\mu,\zeta)$ is governed by eq.~(\ref{RGE}).  A systematic treatment of  this case has been provided in ref.~\cite{Scimemi:2018xaf}, and in our results  we have implemented the {\it optimal} solution suggested in that work. Summarizing the main points:
The solution of eq.~(\ref{RGE}) is in principle path independent, when the anomalous dimensions are known to all orders. This means that the evolution is effectively provided by an evolution potential: in the plane $(\mu,\zeta)$ one can identify null-evolution curves corresponding to equipotential lines and the true evolution occurs only between jet functions belonging to different equipotential lines. 
When  the perturbative expansion of the anomalous dimension is truncated, it is possible to recover a path-independent  result through e.g.~the improved-$\gamma$ scheme of ref.~\cite{Scimemi:2018xaf}, which only affects terms in the perturbative expansion beyond the order that one is working at.

At this point we are left to choose an initial equipotential line $\zeta_\mu(\vecb{b})$, which is known as the $\zeta$-prescription. A special line is provided  by the saddle point of the evolution potential. This line exists for all values of $\vecb{b}$ (at least for $b_T< 1/\Lambda_{QCD}$) and covers all the ranges on $\mu$ and $\zeta$, providing the optimal solution
%%%
\begin{align}
J_q(\vecb{b};\mu,\zeta_\mu(\vecb{b}))=J_{q}(\vecb{b}).
\end{align}
%%%
Explicitly, at two-loop order
%%%
\begin{align}
  \textbf{l}_{\zeta_\mu} \equiv \ln \frac{\mu^2}{\zeta_\mu} &= \frac{1}{2}L_\mu-\frac{3}{2} + 
  a_s\bigg[\frac{11C_A-4n_fT_F}{36}L_\mu^2
  \\ & \quad
  + C_F\Big(-\frac{3}{4}+\pi^2-12\zeta_3\Big) + 
  C_A\Big(\frac{649}{108}-\frac{17\pi^2}{12}+\frac{19}{2}\zeta_3\Big)+
  n_fT_F\Big(-\frac{53}{27}+\frac{\pi^2}{3}\Big)\bigg]
  \, .\nn
\end{align}
%%%

The evolution of the optimal distribution to a generic set of scales $(\mu,\zeta)$ is then simply given by 
%%%
\begin{align}\label{th:evolution}
J_q(\vecb{b};\mu,\zeta)=J_q(\vecb b) U^q_R[\vecb{b}; (\mu,\zeta),(\mu_0,\zeta_{\mu_0}(\vecb b))],
\end{align}
%%%
where  $(\mu_0,\zeta_{\mu_0}(\vecb b))$ is a point on the special line and $U^q_R$ is the TMD evolution factor 
%%%
\begin{align} \label{eq:R}
U^q_R[\vecb b;(\mu_1,\zeta_1),(\mu_2,\zeta_2)]=\exp\[\int_P \( \gamma_q(\mu,\zeta) \frac{\df \mu}{\mu}-\mathcal{D}_q(\mu,\vecb b)\frac{\df \zeta}{\zeta}\)\].
\end{align}
%%%
Choosing the simplest possible line which connects the initial and final point of the evolution in the  improved-$\gamma$ scheme, eq.~(\ref{eq:R}) reduces 
to\footnote{The scales in the argument of $U_R^q$ are ordered according to the convention of \cite{Scimemi:2018xaf}.} 
%%%
\begin{align}\label{th:evolution_our}
U_R^q[\vecb b; (\mu,\zeta),(\mu,\zeta_\mu(\vecb b)) ]=U_R^q[\vecb b;(\mu,\zeta)]=\(\frac{\zeta}{\zeta_\mu(\vecb b)}\)^{-\mathcal{D}_q(\mu,\vecb b)},
\end{align}
%%%
which is convenient for numerical calculations.

The rapidity anomalous dimension $\mathcal{D}_q$ has a nonperturbative part, which is independent of other nonperturbative inputs of the jet distribution and should be estimated by itself.
The $\zeta$-prescription (unlike e.g.~the $b^*$-prescription) allows this separation theoretically. At the moment, the only extraction of the nonperturbative part of the evolution factor from data within this prescription has been carried out in ref.~\cite{Bertone:2019nxa}, so that in our phenomenological analysis we use their parametrization for the nonperturbative contribution to the rapidity anomalous dimension,
%%%
\begin{align}\label{model:rad}
\mathcal{D}_q(\mu,\vecb b)=\mathcal{D}_q^{\text{res}}\(\mu,b^*(\vecb b)\)+g(\vecb b).
\end{align}
%%%
Here $\mathcal{D}_q^\text{res}$ is the resummed perturbative part of $\mathcal{D}_q$, and
%%%
\begin{align} \label{eq:c0def}
b^*(\vecb b)=\sqrt{\frac{b_T^2 B_{\text{NP}}^2}{b_T^2+B_{\text{NP}}^2}}
\,, \qquad
g(\vecb b)=c_0\, b_T\, b^*(\vecb b)
\,,\end{align}
%%%
where the constants $B_{\rm NP}$ and $c_0$ parametrize the nonperturbative effects.
The perturbative expansion of the resummed rapidity anomalous dimension $\mathcal{D}_q^{\text{res}}$ is 
%%%
\begin{align}\label{model:d_resum}
\mathcal{D}_q^{\text{res}}\(\mu,\vecb b\)=\sum_{n=0}^\infty a_s^n(\mu)  \mathcal{D}_q^{\text{res}[n]}(X)\
\end{align}
%%%
where $X=\beta_0 a_s(\mu)\ln(\mu^2 b_T^2 e^{2\gamma_E}/4)$, $\beta_0$ is the leading coefficients of the QCD beta function and $a_s=g^2/(4\pi)^2$. The leading term reads
%%%
\begin{align}\label{model:d0}
 \mathcal{D}_q^{\text{res}[0]}(X)=-\frac{\Gamma^\zero_q}{2\beta_0}\ln(1-X),
\end{align}
%%%
and we have used this expansion up to third order in $a_s$, which incorporates the four-loop anomalous dimension.
The complete expression up this order can be found in ref.~\cite{Echevarria:2012pw,Scimemi:2018xaf}.
The unresummed expression for the rapidity anomalous dimension is reported in \eq{DAnomDim}.

%===============================================================================
\subsection{Numerical implementation of evolution}
\label{sec:jet_ren}
%===============================================================================

We use \Artemide to run the double scale evolution from the initial scale of the TMD jet function/PDF
%%%
\begin{align} \label{eq:inScale}
(\mu_0,\zeta_0)= \Big(\frac{2e^{-\gamma_E}}{b_T} + 2\mbox{ GeV}\,,\zeta_{\mu_0} \Big)\, ,
\end{align}
%%%
where $\mu_0$ is frozen at 2 GeV to avoid the Landau pole and  $(\mu_0,\zeta_0)$ belongs to the special line, to the hard scale
%%%
\begin{align} \label{eq:finScale}
 (\mu_H,\zeta_H) = 
 \begin{cases}
 ( \sqrt{s}, s ) \qquad &e^+e^- \\
 ( Q, Q^2 ) \qquad &\mbox{SIDIS}
 \end{cases} \
\end{align}
%%%

Since the rapidity resummation is the dominant source of uncertainty and to consistently use the nonperturbative parameters extracted in ref.~\cite{Bertone:2019nxa}, we will always use the highest known order in the evolution, even though the jet function for generic $R$ is only calculated at one-loop order.
The nonperturbative parameters of the evolution kernel in eqs.~\eqref{model:rad} and \eqref{eq:c0def} are set to 
%%%
\begin{align}
B_{NP}=2.5 \text{ GeV}^{-1}
\,, \qquad
c_0 = 0.037
\,.\end{align}
%%%

%%%%%%%%%%%%%%%%%%%%%%%%%%%%%%%%%%%%%%%%%%%%%%%%%%%%%%%%%%%%%%%%%%%%%%%%%%%%%%%%
\section{Quark jet function for large $R$ at two loops}
\label{sec:jet_two}
%%%%%%%%%%%%%%%%%%%%%%%%%%%%%%%%%%%%%%%%%%%%%%%%%%%%%%%%%%%%%%%%%%%%%%%%%%%%%%%%

%%%
\begin{figure*}[tb]
  \centering
  \includegraphics[width=0.45\textwidth]{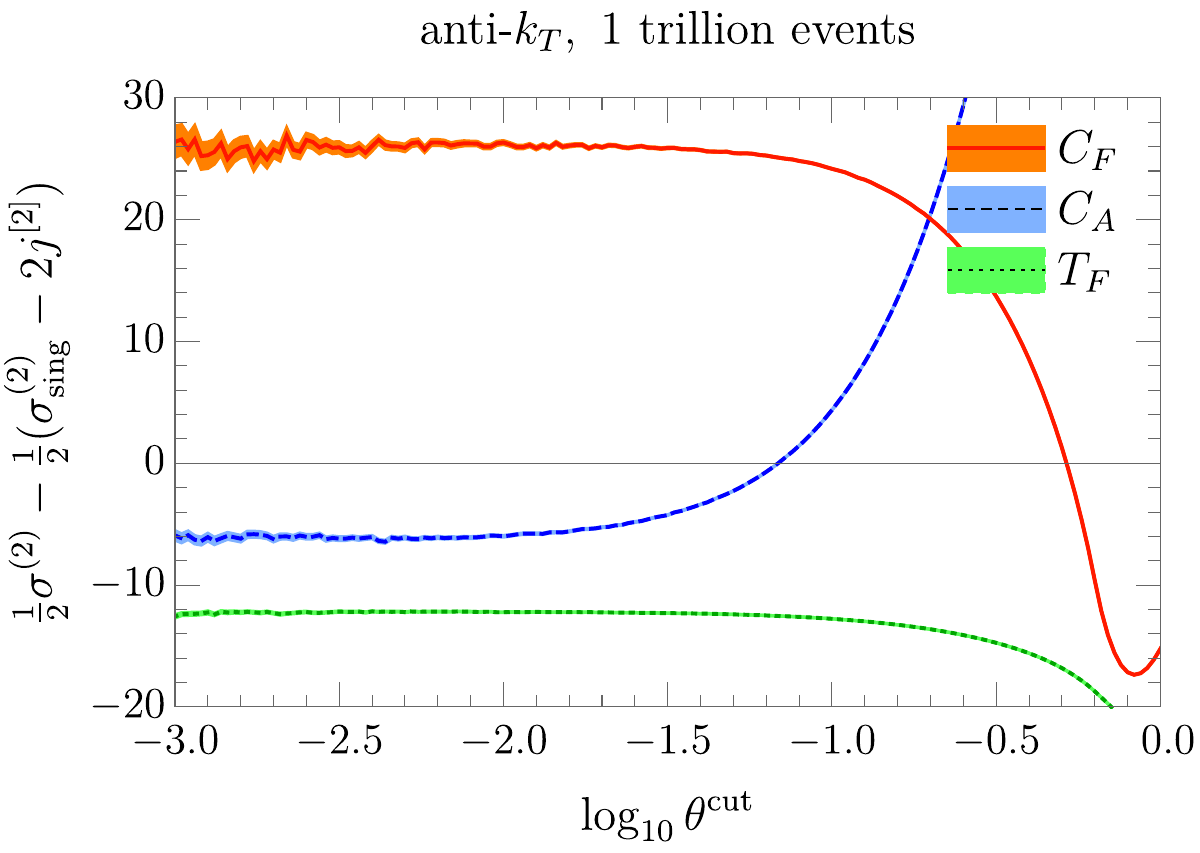} \quad
  \includegraphics[width=0.45\textwidth]{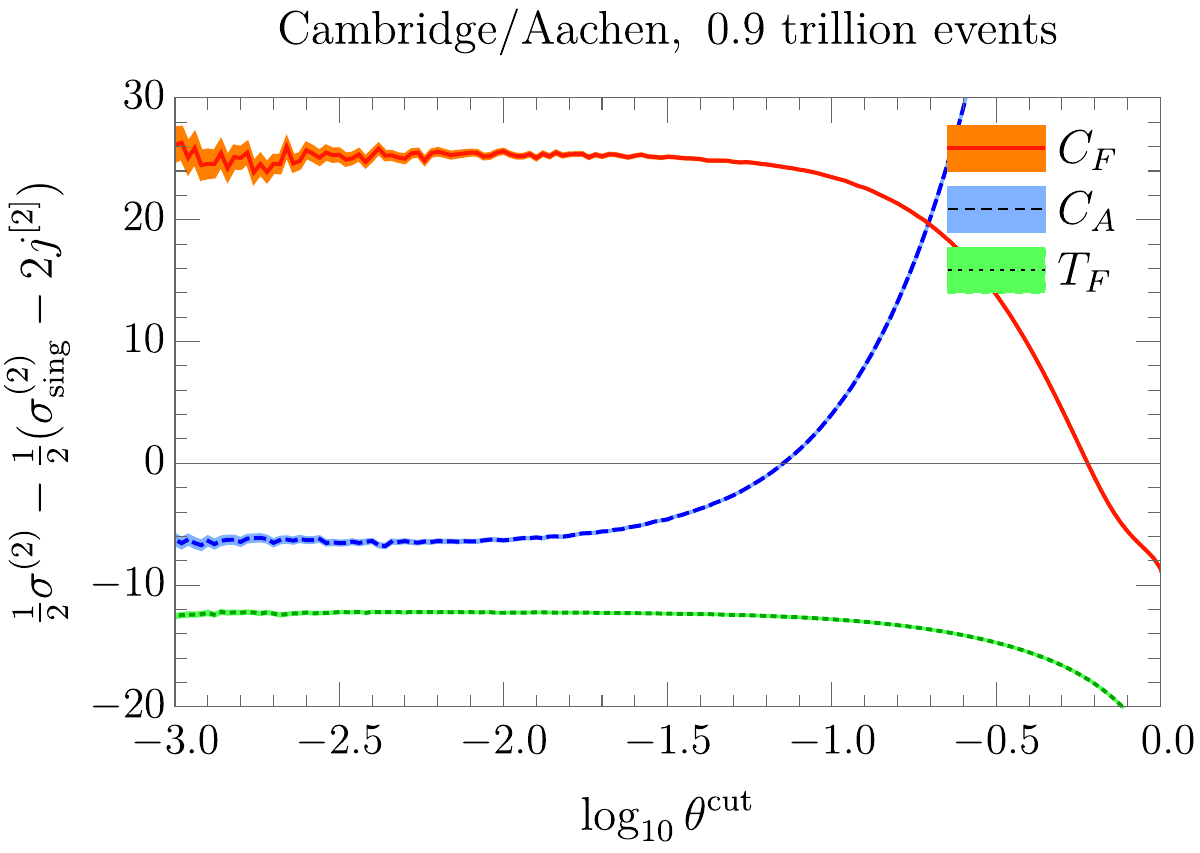} \\
  \includegraphics[width=0.45\textwidth]{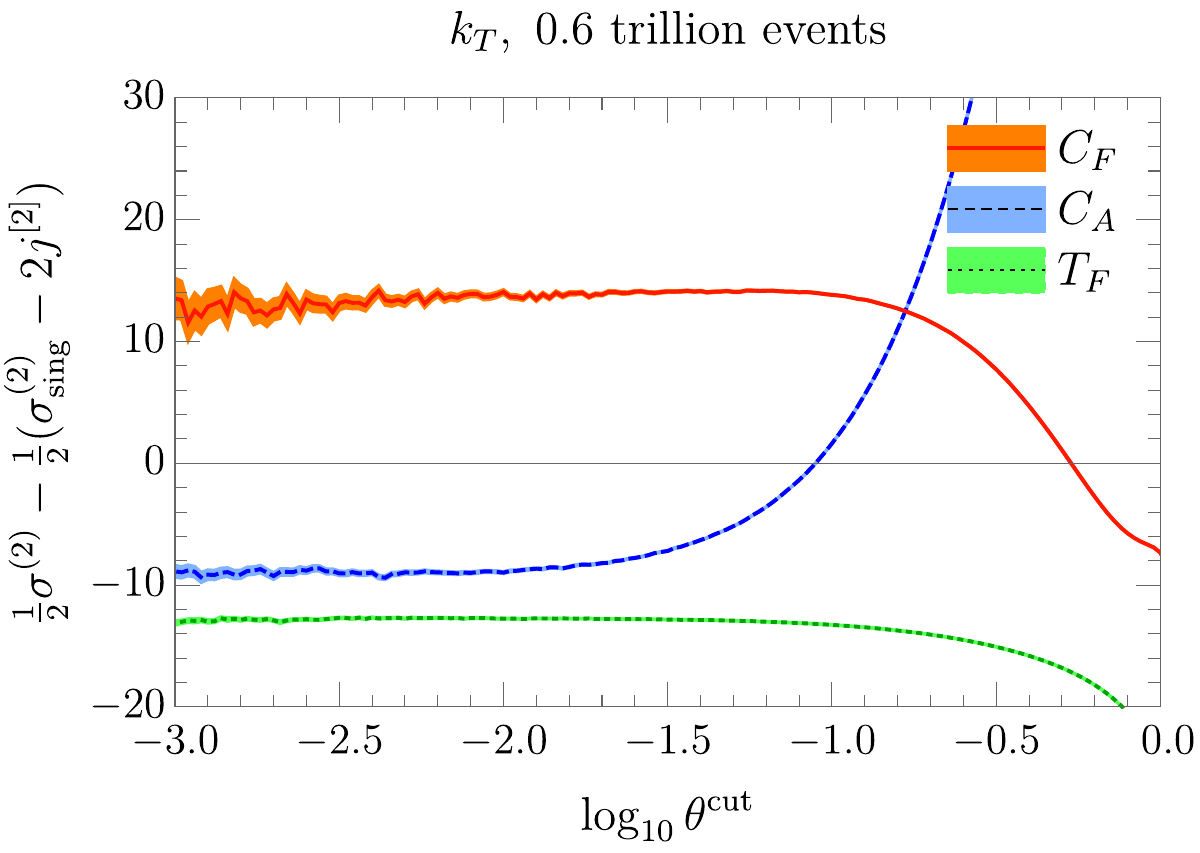}
  \caption{The difference between the $\ord{\al_s^2}$ contribution to $e^+e^-$ cross section with a cut on the angular decorrelation  $\theta \leq \theta^\cut$, obtained from \Eventtwo and from our factorization theorem. The panels correspond to the ($e^+e^-$ version of) anti-$k_T$, Cambridge/Aachen and $k_T$ jet algorithm, and the curves correspond to the different color structures, see \eq{j2_color}. The uncertainty bands indicate the statistical uncertainty. The missing two-loop constant in the quark jet function is the value of the plateau at small $\theta^\cut$.}
\label{fig:event2}
\end{figure*}
%%%

As we will see in our numerical analysis, the large-$R$ limit captures the dominant part of the perturbative corrections. This justifies focusing on the quark jet function in the large-$R$ limit, ${\mathscr{J}}_q^{\WTA}$, which is completely determined at two loops by known anomalous dimensions, except for a constant $j^{[2]}$. Explicitly, 
%%%
\begin{align}
  \nn {\mathscr{J}}_q^{\two\, ,\WTA}(\vecb{b},\mu,\zeta) &= C_F\Bigg\{
  C_F\bigg[
  \frac{1}{2}L_\mu^4-(3+2\textbf{l}_\zeta)L_\mu^3
  + \Big(2\textbf{l}_\zeta^2+6\textbf{l}_\zeta-\frac{5}{2}+6\ln 2 + \frac{5\pi^2}{6}\Big)L_\mu^2\\ \nn
  &\quad +\bigg(\Big(14-12\ln 2-\frac{5\pi^2}{3}\Big)\textbf{l}_\zeta+\frac{45}{2}-18\ln 2 - \frac{9\pi^2}{2}+24\zeta_3\bigg)L_\mu
  \bigg]\\ \nn
  &\quad +C_A\bigg[-\frac{22}{9}L_\mu^3+
  \Big(\frac{11}{3}\textbf{l}_\zeta-\frac{35}{18}+\frac{\pi^2}{3}\Big)L_\mu^2
  +\Big(\frac{404}{27}-14\zeta_3\Big)\textbf{l}_\zeta\\ \nn
  &\quad +\bigg(\Big(\frac{134}{9}-\frac{2\pi^2}{3}\Big)\textbf{l}_\zeta+\frac{57}{2}-22\ln 2 -\frac{11\pi^2}{9}-12\zeta_3\bigg)L_\mu\bigg]
  +n_fT_F\bigg[
  \frac{8}{9}L_\mu^3 \\& \quad + \Big(\frac{2}{9}-\frac{4}{3}\textbf{l}_\zeta\Big)L_\mu^2-\frac{112}{27}\textbf{l}_\zeta
  +\Big(-\frac{40}{9}\textbf{l}_\zeta-10+8\ln 2+\frac{4\pi^2}{9}\Big)L_\mu\bigg]\Bigg\} +j^\two\, .
\end{align}
%%%
We extract this constant using the \Eventtwo generator~\cite{Catani:1996vz}, which we run with $n_f=5$ and an infrared cutoff $\rho = 10^{-12}$, generating about a trillion events. Specifically, we consider the difference at $\ord{a_s^2}$ between the the cross section with a cut on the angular decorrelation $\theta \leq \theta^\cut$ obtained from \Eventtwo and our factorization theorem, extracting the overall factor of $a_s^2 = \al_s^2/(4\pi)^2$. This is shown in \fig{event2}, where the different panels correspond to the ($e^+e^-$ version of) anti-$k_T$~\cite{Cacciari:2008gp}, Cambridge/Aachen~\cite{Dokshitzer:1997in,Wobisch:1998wt} and $k_T$~\cite{Catani:1991hj} jet algorithm. The different curves in each panel correspond to the $C_F^2$, $C_FC_A$ and $C_F T_F$ color structure, with the bands indicating the statistical uncertainty. From varying the infrared cutoff we conclude that the cross section obtained from \Eventtwo can be trusted for $\log_{10} \theta^\cut > -3$, corresponding to the plotted range.
 
The clear plateau at small values for $\theta^\cut$ shows that our factorization theorem predicts the singular part of the cross section correctly. The value of the plateau corresponds to the missing two-loop constant $j^{[2]}$ (the overall factor of $1/2$ was chosen to cancel the factor of 2 from the two jet functions in the factorization theorem). The decomposition of $j^{[2]}$ in terms of the $C_F^2$, $C_FC_A$ and $C_F T_F$
color structures is given by
%%%
\begin{align} \label{eq:j2_color}
  j^{[2]} = j_{C_F}^{[2]} + j_{C_A}^{[2]} + \frac{n_f}{5} j_{T_F}^{[2]}
\,,\end{align}
%%%
i.e.~the color structures are inside the constants. 
We extracted the result by fitting the plateau to a constant, assuming $n_f=5$, and the generalization to arbitrary number of flavors only involves rescaling $j^\two_{T_F}$. The best range for this fit is not a priori clear, since we have no control over the power corrections, corresponding to contributions to the cross section not included in our factorization theorem. These become more relevant as $\theta^\cut$ increases; on the other hand, lowering $\theta^\cut$ increases the statistical uncertainties. We choose to consider the fit range $-3 \leq \log_{10} \theta^\cut \leq \log_{10} \theta^\cut_{\rm max}$, where we vary $\log_{10} \theta^\cut_{\rm max}$ between $-2.9$ and $-2$ in steps of 0.02 (this corresponds to the size of our binning). We perform a different fit in each window, including the uncertainty from the \Eventtwo integration. We take the lowest and highest value obtained in this way as the error, and their average as the central value, leading to
%The result was extracted for $n_f=5$, and the generalization to arbitrary number of flavors only involves rescaling $j^\two_{T_F}$.
%While the power corrections, corresponding to contributions to the cross section not included our factorization theorem, decrease as $\theta^\cut$ becomes smaller, the statistical uncertainties increase. This motivates us to fit the plateau to a constant, considering as fit range $-3 \leq \log_{10} \theta^\cut \leq \log_{10} \theta^\cut_{\rm max}$, where we vary $\log_{10} \theta^\cut_{\rm max}$ between $-2.9$ and $-2$. We take the spread between the values obtained in this manner as an estimate of our uncertainty, finding
%%%
\begin{align} \label{eq:two_loop_consts}
  \text{anti-}k_T\!:\ & j_{C_F}^{[2]} = 25.3 \pm 0.6 \,, &  j_{C_A}^{[2]} &= -6.3 \pm 0.2 \,, & j_{T_F}^{[2]} &= -12.5 \pm 0.3\,, \nn \\
  C/A\!:\ & j_{C_F}^{[2]} = 24.5 \pm 0.6 \,, &  j_{C_A}^{[2]} &= -6.7 \pm 0.2 \,, & j_{T_F}^{[2]} &= -12.5 \pm 0.2\,, \nn \\
   k_T\!:\ & j_{C_F}^{[2]} = 12.2 \pm 1.1\,, &  j_{C_A}^{[2]} &= -9.3 \pm 0.2\,, & j_{T_F}^{[2]} &= -13.0 \pm 0.3\,.
\end{align}
%%%
While these constants are remarkably similar for anti-$k_T$ and Cambridge/Aachen, they differ substantially for $k_T$. 

%%%%%%%%%%%%%%%%%%%%%%%%%%%%%%%%%%%%%%%%%%%%%%%%%%%%%%%%%%%%%%%%%%%%%%%%%%%%%%%%
\section{Results}
\label{sec:results}
%%%%%%%%%%%%%%%%%%%%%%%%%%%%%%%%%%%%%%%%%%%%%%%%%%%%%%%%%%%%%%%%%%%%%%%%%%%%%%%%

The region of interest for TMDs is small $q_T$, for which the regimes $\theta \sim R$ and $\theta \ll R$ are most relevant. This leads us to exclusively focus on the WTA axis, which is well behaved in the large-$R$ limit. We start by considering the transverse momentum decorrelation in $e^+e^-$ collisions, obtaining numerical predictions for the Belle II and LEP experiments. We use $e^+e^-$ to test the perturbative convergence, and explore the dependence on the jet radius $R$ and cut on the jet energy fraction $z$. In the case of SIDIS we provide numerical predictions for HERA and the EIC, and investigate the sensitivity of our cross section to nonperturbative effects.

In our numerical implementation we build on the \Artemide code~\cite{Scimemi:2017etj,Scimemi:2018xaf} to obtain resummed predictions for TMD cross sections. 
The original version of \Artemide~\cite{artemide} provides cross sections for Drell-Yan and SIDIS with fragmentation into hadrons. However, its modular structure allowed us to extend it to processes involving jets with a modest amount of modification. Specifically, we have added $e^+e^- \to$ dijet and jet-SIDIS high-level modules, and a jet TMD low-level module that provides our perturbative input for the quark jet functions in $\vecb{b}$-space at the initial scale.
 
%===============================================================================
\subsection{Momentum decorrelation in $e^+e^-$ collisions} \label{sec:ee}
%===============================================================================

%%%
\begin{figure*}[tb]
  \centering
  \includegraphics[width=0.48\textwidth]{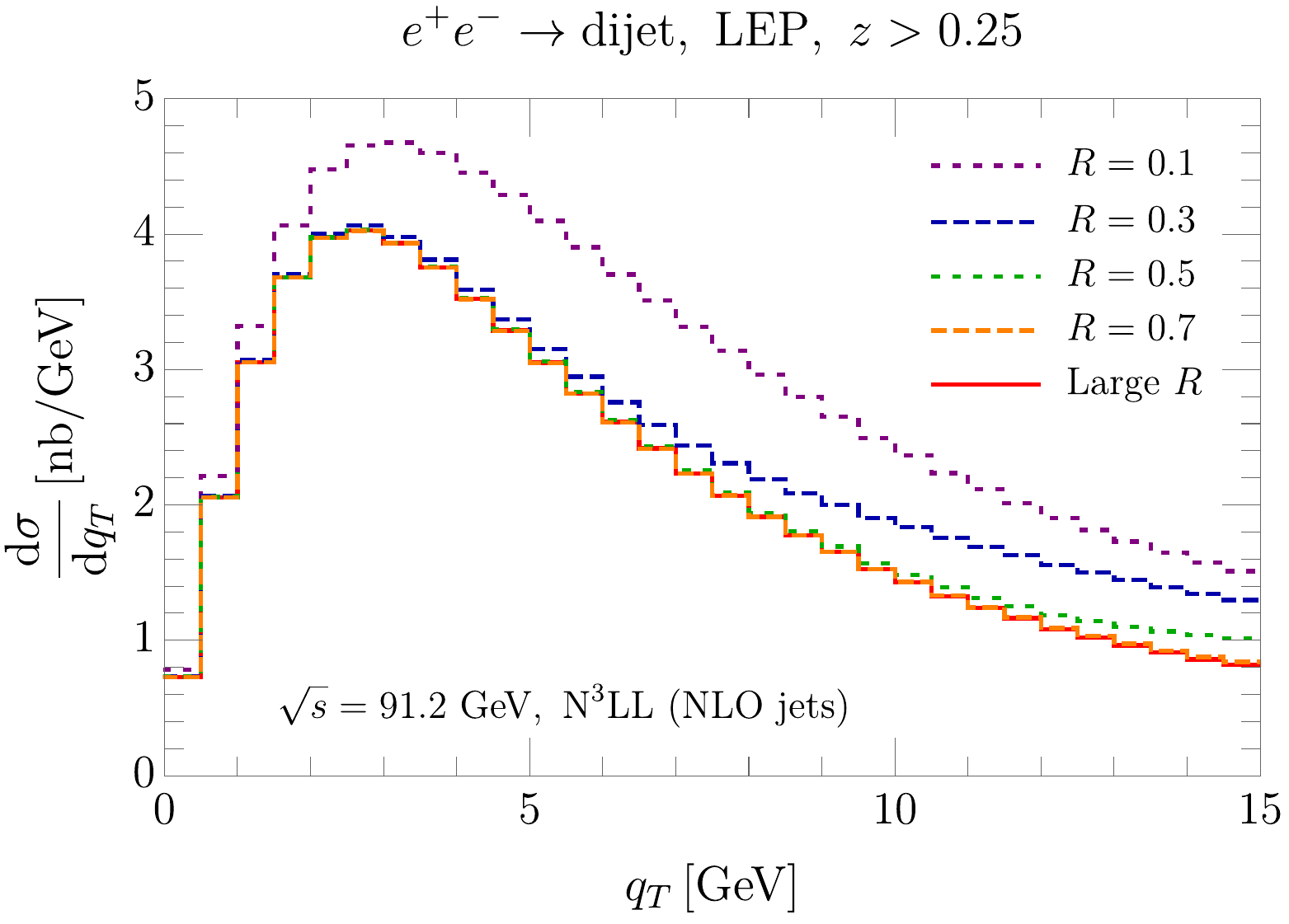} \quad
  \includegraphics[width=0.48\textwidth]{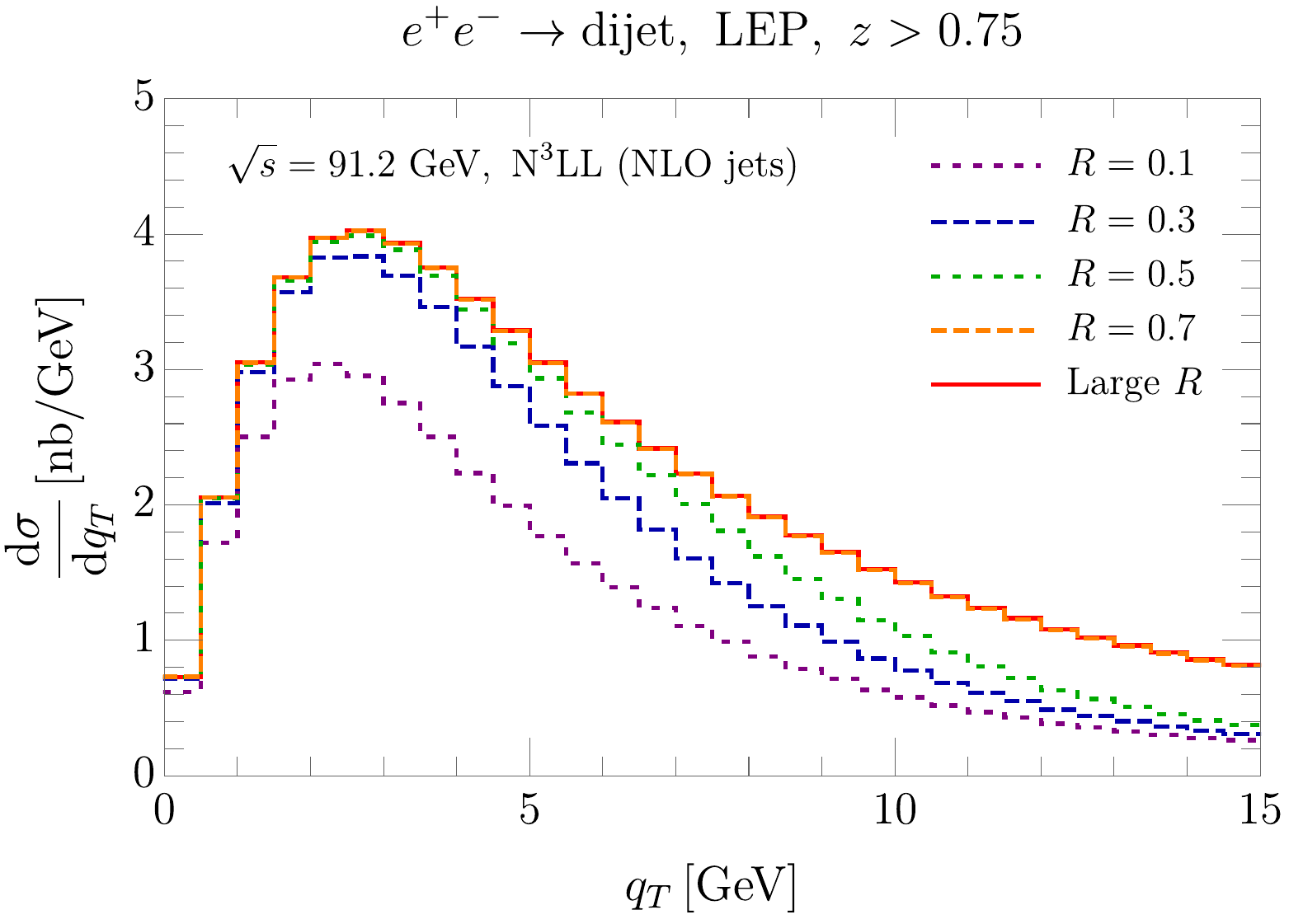} \\
  \caption{Dependence of the cross section differential in the transverse momentum decorrelation  on the jet radius parameter $R$, for cuts on jet energy fraction $z>0.25$ (left) and $z>0.75$ (right). We use the NLO jet function computed in the regime $R \sim \theta$, and show the large-$R$ result (red solid) for comparison.}
\label{fig:varyR}
\end{figure*}
%%%

In our analysis of the $e^+e^-$ cross section, differential in the transverse momentum decorrelation, we consider two experiments:
\begin{itemize}
\item Belle II: $\sqrt{s} = 10.52$ GeV, 4 quark flavors. 
\item LEP: $\sqrt{s} = 91.1876$ GeV, 5 quark flavors.
\end{itemize}
We account for both the photon and $Z$-boson contribution, and restrict the plotted $q_T$ range to a region where the power corrections to the factorization theorem can be neglected. In the Belle analysis we omit $b$-jets, since we do not include quark mass effects in our calculation of the jet function. (Experimentally, these are of course relatively easy to distinguish from light quark jets.)

We start our analysis by studying the dependence on the jet radius parameter $R$ in \fig{varyR} for LEP. The cross section is shown for various jet radii, ranging from $R=0.1$ to $0.7$, using the factorization formulae for $\theta \sim R$ in \sec{fac1}. We consider two representative cuts on the jet energy fraction: $z>0.25$ (left panel) and $z>0.75$ (right panel). For comparison we also show the large-$R$ limit, discussed in \sec{fac34_wta}. We use the one-loop jet function (since we only have the one-loop result for $\theta \sim R$), but include the hard function at two-loop order and perform the resummation at N$^3$LL accuracy.

As expected, as $R$ increases the results approach the $R\to\infty$ limit. In both cases, the cross section for $R=0.7$ is indistinguishable from the large-$R$ result, and for $z > 0.25$ the difference is even minimal for $R=0.5$. This means that in the factorization in \eq{J_wta} the power corrections  $\ord{\theta/R}\sim \ord{b_T^2 E^2 \mathcal{R}^2}$ have a limited impact even for $\theta \lesssim R$. This observation will be used in the rest of our analysis, to justify including the two-loop jet function in the large-$R$ limit, as this will capture the dominant two-loop contribution. Explicitly, we will combine results according to 
%%%
\begin{equation} \label{eq:combineLargeR}
  \bigg(\frac{\df\sigma}{\df q_T}\bigg)^{\rm N^3LL}  = 
   \bigg(\frac{\df\sigma}{\df q_T}\bigg)^{\rm NLO} + 
    \bigg(\frac{\df\sigma}{\df q_T}\bigg)^{\rm NNLO}_{\cR \to \infty} -
     \bigg(\frac{\df\sigma}{\df q_T}\bigg)^{\rm NLO}_{\cR \to \infty}\, ,
\end{equation}
%%%
where NLO and NNLO indicate the order of the jet function. In each term we use the NNLO hard function and include the resummation at N$^3$LL accuracy.
The above approximation contains all large logarithms of $\theta$ (or equivalently, $q_T$) at  N$^3$LL accuracy. It reduces to NNLL accuracy for $\theta \sim R \ll 1$, since it misses some $\ord{\theta/R}$ corrections. We have shown that their effect is small, except in the tail region.

%%%
\begin{figure*}[tb]
  \centering
  \includegraphics[width=0.48\textwidth]{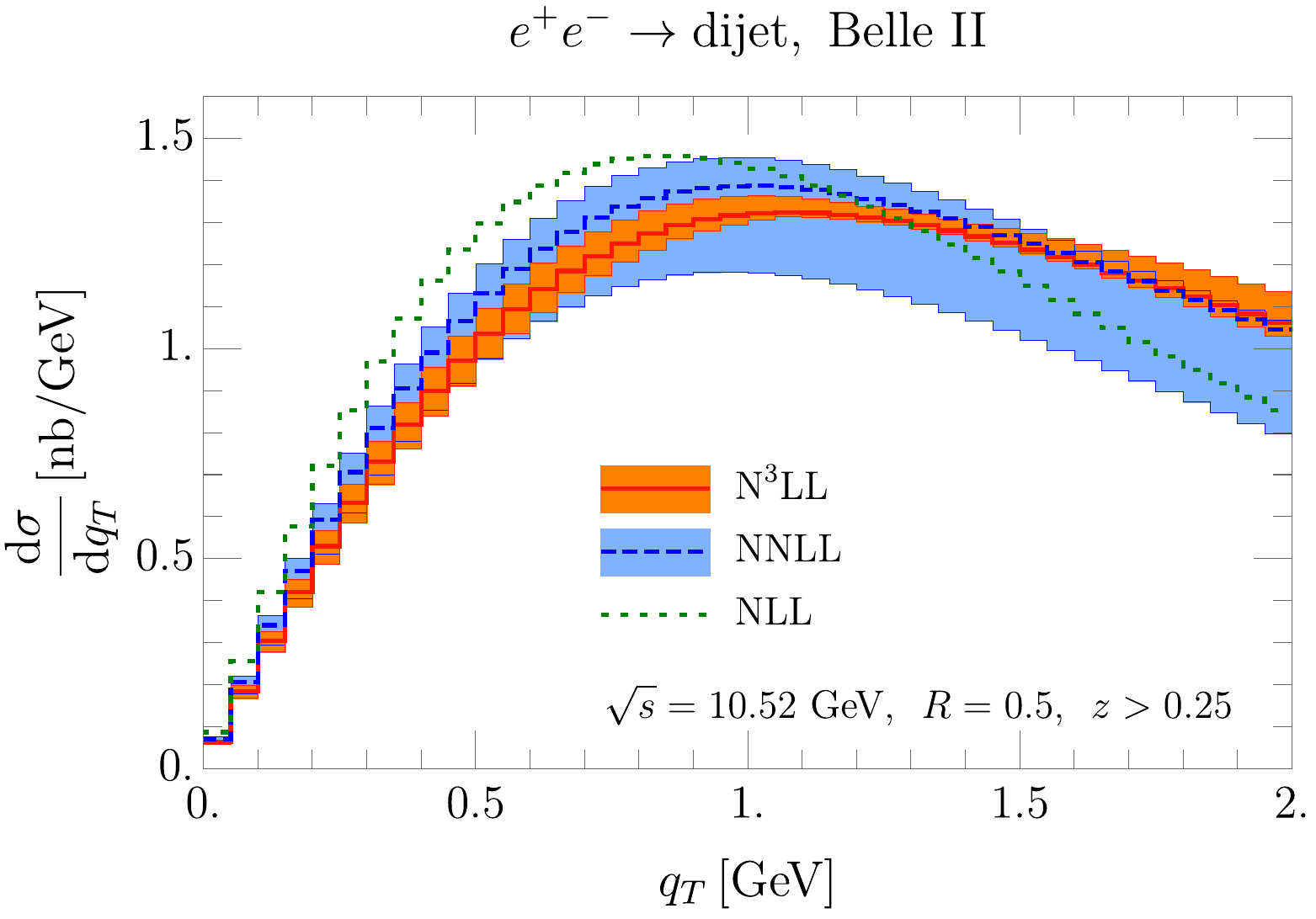} \quad
  \includegraphics[width=0.48\textwidth]{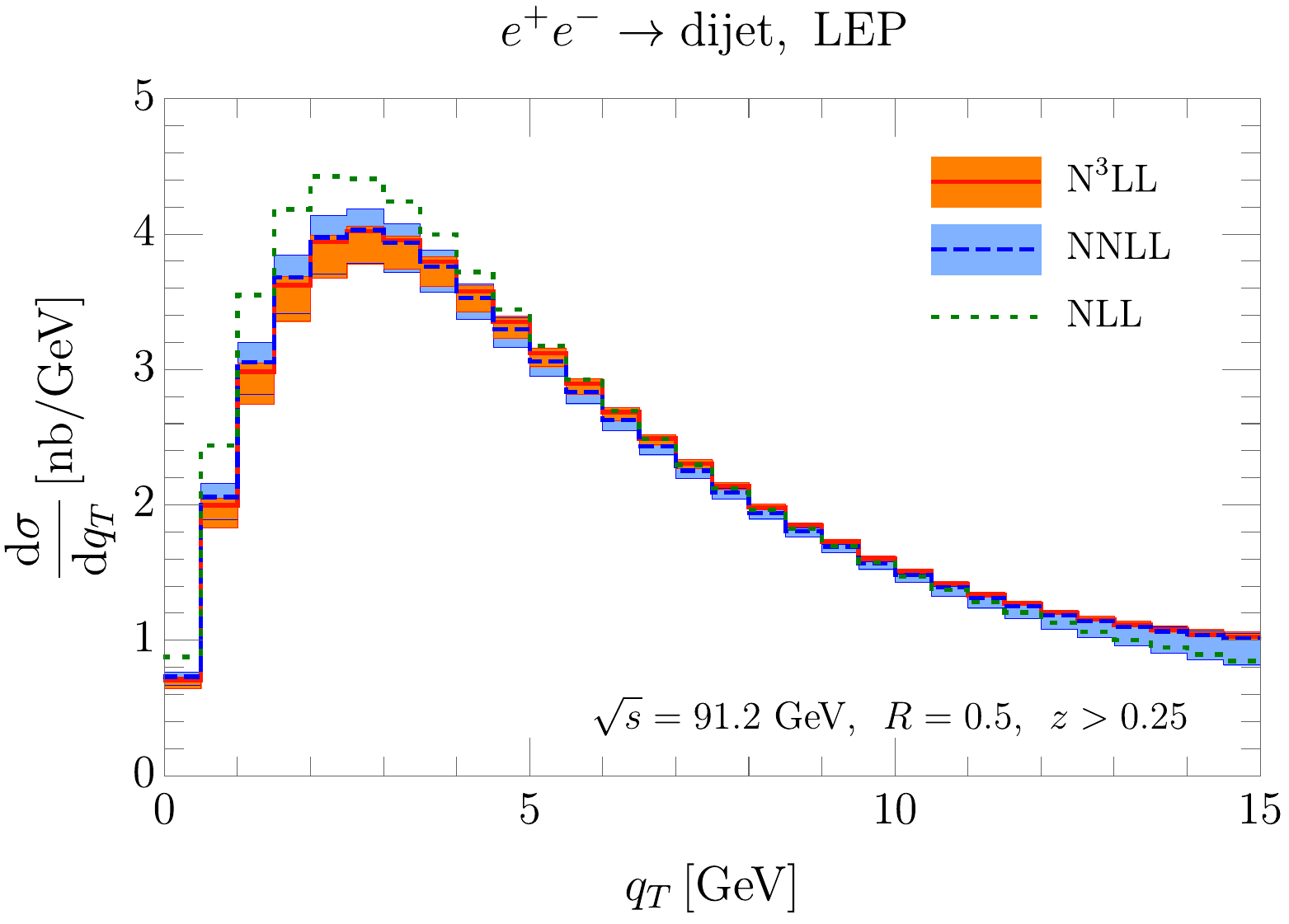} \\
  \caption{Perturbative convergence of the cross section differential in transverse momentum decorrelation, for Belle II (left) and LEP (right), for jet radius $R=0.5$ and jet energy fraction $z>0.25$. The N$^3$LL result is obtained with the prescription in \eq{combineLargeR}. The bands encode the perturbative uncertainty, as described in the text.}
\label{fig:convergencePlot}
\end{figure*}
%%%

Next we study the perturbative convergence of the TMD cross section in \fig{convergencePlot}. We take $R=0.5$, $z>0.25$ and show results for the cross section for Belle II (left panel) and LEP (right panel) at NLL, NNLL and N$^3$LL. The ingredients that enter in the various perturbative orders are summarized in table~\ref{tab:ordersLog}. The perturbative uncertainty is estimated by varying the scales $\mu_i$ in \eqs{inScale}{finScale} up and down by a factor $2$ around their central value and taking the envelope. The band obtained by this procedure at NLL is artificially small and not shown.
As expected, the N$^3$LL correction is small compared to the NNLL one, and the uncertainty bands overlap and are reduced at higher order.

%%%
\begin{figure*}[tb]
  \centering
  \includegraphics[width=0.48\textwidth]{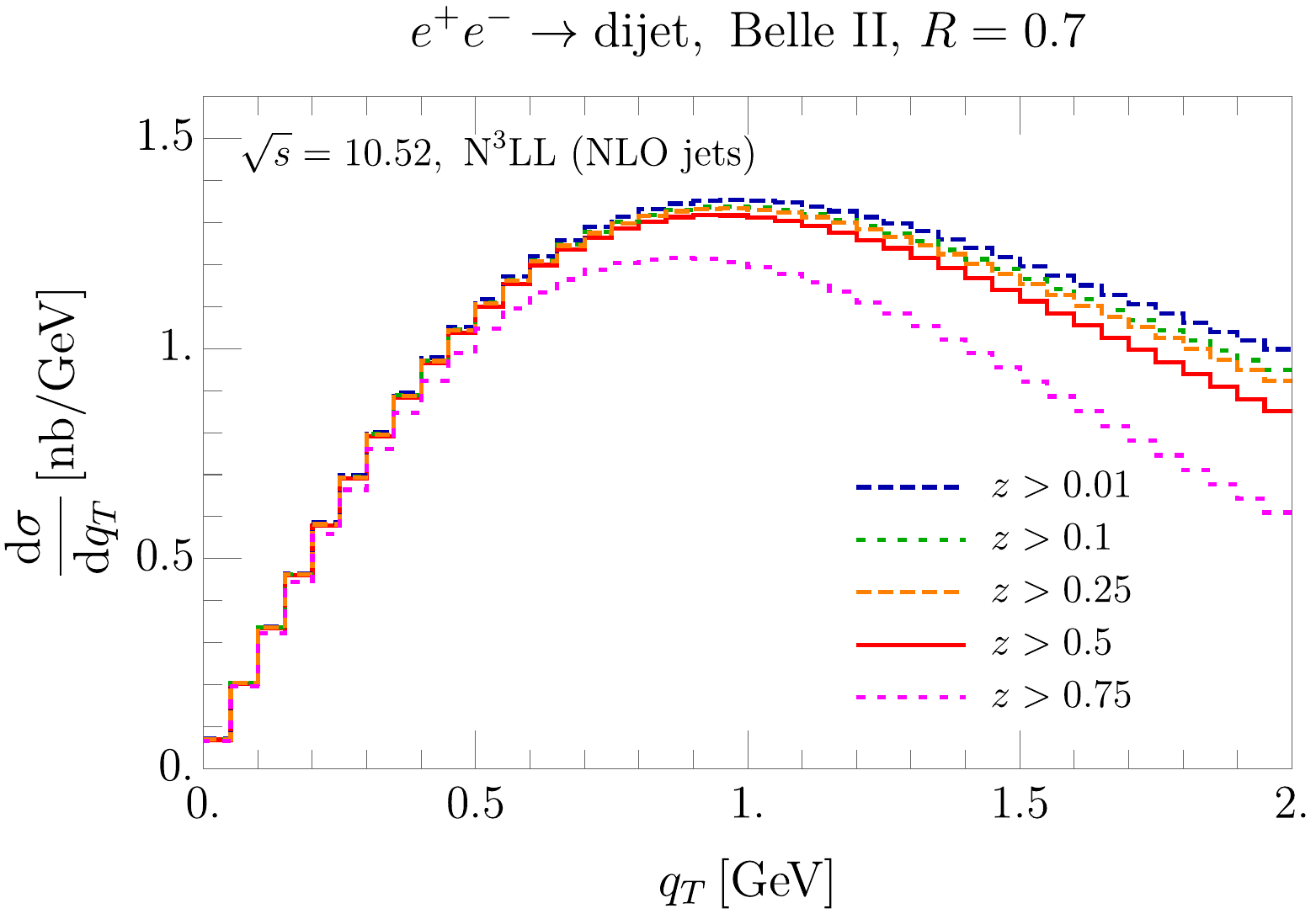} \quad
  \includegraphics[width=0.48\textwidth]{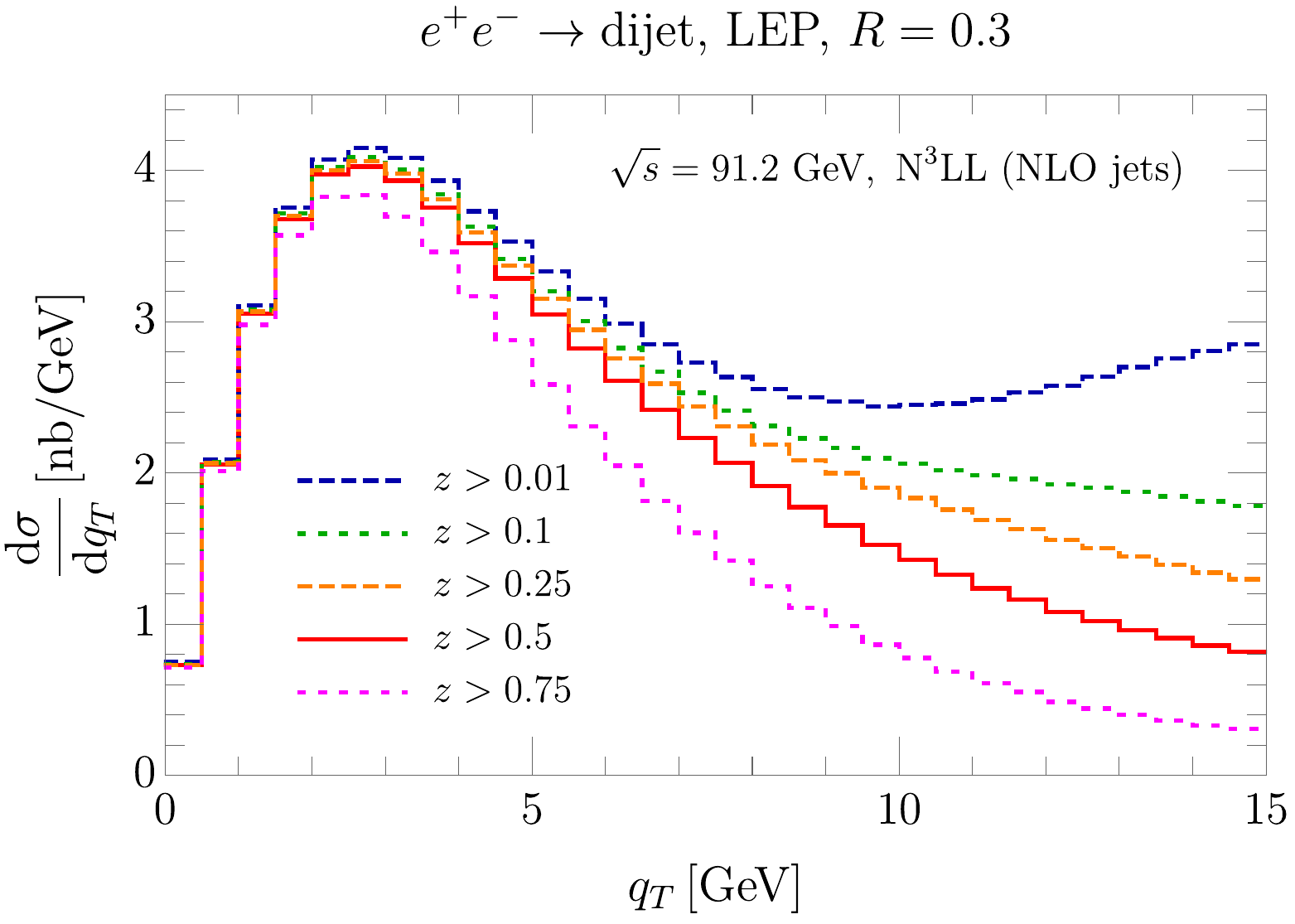} \\
  \caption{Dependence of the transverse momentum decorrelation distribution on the cut on jet energy fraction $z$, for Belle II with $R=0.7$ (left panel) and LEP with $R=0.3$ (right panel). The dependence on this cut is larger for smaller $R$, as discussed in the text. In both cases, the results for $z > 0.5$ (solid red curve) exactly coincide with the large-$R$ limit, see footnote.}
\label{fig:varyZ}
\end{figure*}
%%%

In \fig{varyZ} we investigate the dependence of the cross section on the cut  on the jet energy fraction $z>z_\cut$ for a fixed value of the jet radius, which provides a complementary picture to \fig{varyR}. We show results for Belle II with $R=0.7$ (left panel) and LEP with $R=0.3$ (right panel), imposing $z>z_{\rm{cut}}$ and varying $z_{\rm{cut}} = 0.01$ to $z_{\rm{cut}} = 0.75$. As in \fig{varyR}, we use NLO jet functions. 
For $R=0.7$ the dependence on the cut on $z$ is relatively mild, which reflects the fact that in the large-$R$ limit the jet function is proportional to $\delta(1-z)$, and thus independent of this cut.
For $R=0.3$ there is a stronger dependence, and at very small (large) values of $z$ the cross section shows unphysical features. This is not surprising, since the cross section diverges as $z_{\rm{cut}} \to 0$ (every single low-energy particle originates a different jet) and has large logarithms of $1-z_{\rm{cut}}$ for $z_{\rm{cut}} \to 1$. 
We found that, regardless of the jet radius, for $z_{\rm{cut}} = 0.5$ the cross section coincides with the large-$R$ result. This is due to a one-loop accident.\footnote{At one loop, the initial quark undergoes a single splitting, see \fig{diagrams}. When integrating over $0.5< z < 1$, each phase-space configuration contributes to the cross section with exactly one jet (either a jet containing two particles or a jet containing the most energetic particle). Due to the WTA recombination prescription, the resulting jet axis is the same in either case, independent of $R$. Thus it must in particular coincide with the large-$R$ limit.} 

%%%
\begin{figure*}[tb]
  \centering
  \includegraphics[width=0.495\textwidth]{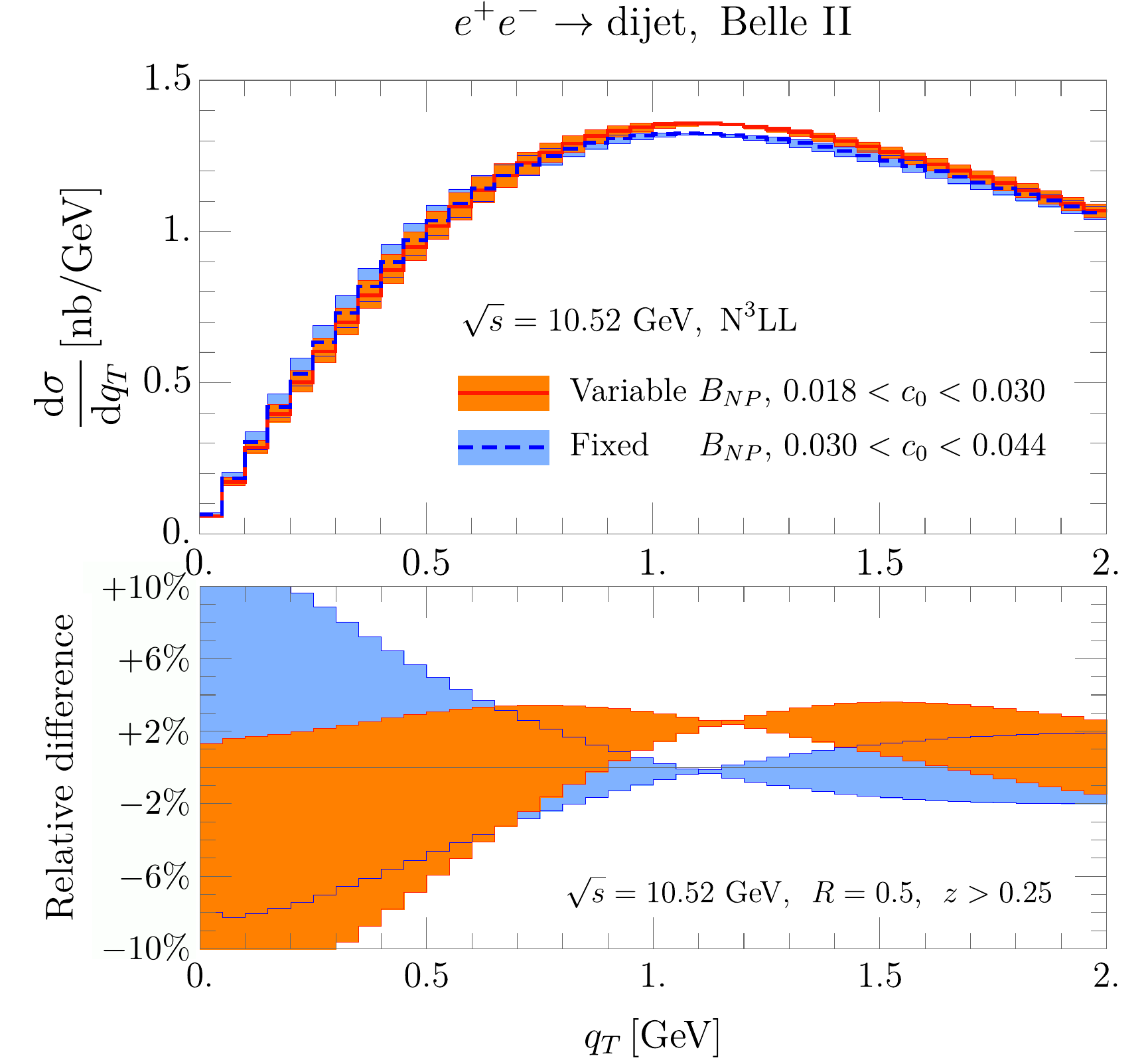}
  \includegraphics[width=0.495\textwidth]{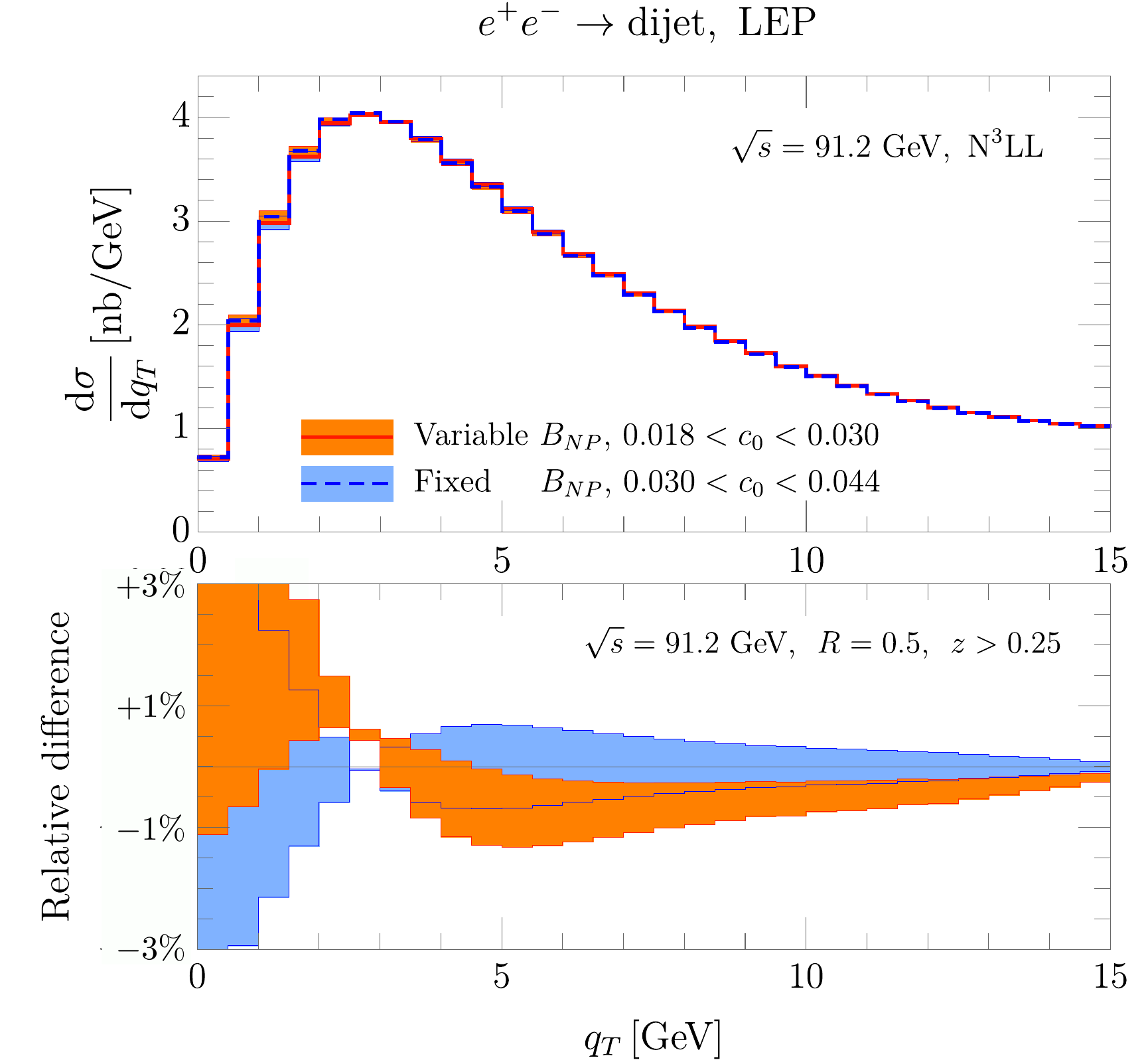}\\	
  \caption{Estimate of the sensitivity of the TMD to nonperturbative effects in the rapidity resummation at Belle II (left) and LEP (right). We vary the parameter $c_0$ in the range of its statistical uncertainty, testing both the fixed and variable $B_{NP}$ schemes of ref.~\cite{Bertone:2019nxa}. Results are obtained with the prescription in \eq{combineLargeR}.}
\label{fig:e+e-varyNP}
\end{figure*}
%%%
As a next step, we study how sensitive these cross sections are to $B_{NP}$ and $c_0$ that parametrize the nonperturbative contribution to the rapidity evolution, see
eqs.~(\ref{model:rad}) and (\ref{eq:c0def}). We considered both the ``fixed $B_{NP}$'' and ``variable $B_{NP}$'' schemes used in the recent fit in ref.~\cite{Bertone:2019nxa}, and varied the parameters within the statistical errors listed in their table 4. In practice, we found that the $B_{NP}$ variation is subdominant, so in \fig{e+e-varyNP} we simply plot variations of $c_0$. As one would expect, the sensitivity to nonperturbative effects is much larger at Belle, commensurate with its smaller center-of-mass energy, and increases at low transverse momenta. The conclusions obtained from the two schemes are compatible with each other. The situation is similar for LEP, though the relative variation is substantially lower (below $1\%$ for most of the range in $q_T$).

Finally, we have investigated the impact of the choice of jet algorithm, specifically, the impact of the different two-loop constants in \eq{two_loop_consts}. We found the difference with respect to anti-$k_T$ to be negligible for Cambridge-Aachen ($< 0.1\% $) and very small for the $k_T$ algorithm ($< 1\% $).

%===============================================================================
\subsection{Transverse momentum dependent distributions in SIDIS}
\label{sec:sidis}
%===============================================================================

In this section we show representative results for TMD measurements with jets in SIDIS, showing results for
\begin{itemize}
\item HERA: $\sqrt{s} = 318$ GeV,
\item EIC: $\sqrt{s} = 100$ GeV.
\end{itemize}
The EIC is a future facility for the study of TMD distributions, and the above value for its center-of-mass energy is an assumption. We take $10 \leq Q \leq 25$ GeV and study the transverse momentum distribution for $q_T \leq 3$ GeV, ensuring that power corrections of order $q_T^2/Q^2$ to the factorization theorem can be neglected. In this kinematic range we expect quark mass effects to be negligible, so we ignore them. We work in the Breit frame, impose a cut on the jet energy fraction $z > 0.25$ and set the jet radius to $R=0.5$. Our $e^+e^-$ analysis in the left panel of \fig{varyR} shows that in this case the large-$R$ approximation works extremely well, so we again include the two-loop, large-$R$ jet function of \sec{jet_two}, using \eq{combineLargeR}.

We use the quark TMD PDFs obtained in ref.~\cite{Bertone:2019nxa}. In this fit the matching of the TMDs onto PDF is incorporated at NNLO, using the NNPDF~3.1 PDFs~\cite{Ball:2017nwa} with $\alpha_s(M_Z) = 0.118$. The additional nonperturbative component of the TMD PDFs is modeled with the ansatz
%%%
\begin{equation} \label{eq:f_NP}
  f_{NP} = \exp \bigg( - \frac{\lambda_1(1-x)+\lambda_2 x + \lambda_3 x(1-x) b_T^2}{\sqrt{1+\lambda_4 x^{\lambda_5}b_T^2}}\bigg) \,,
\end{equation}
%%%
where the values for $\la_i$ were fit in ref.~\cite{Bertone:2019nxa}.

%%%
\begin{figure*}[tb]
  \centering
  \includegraphics[width=0.49\textwidth]{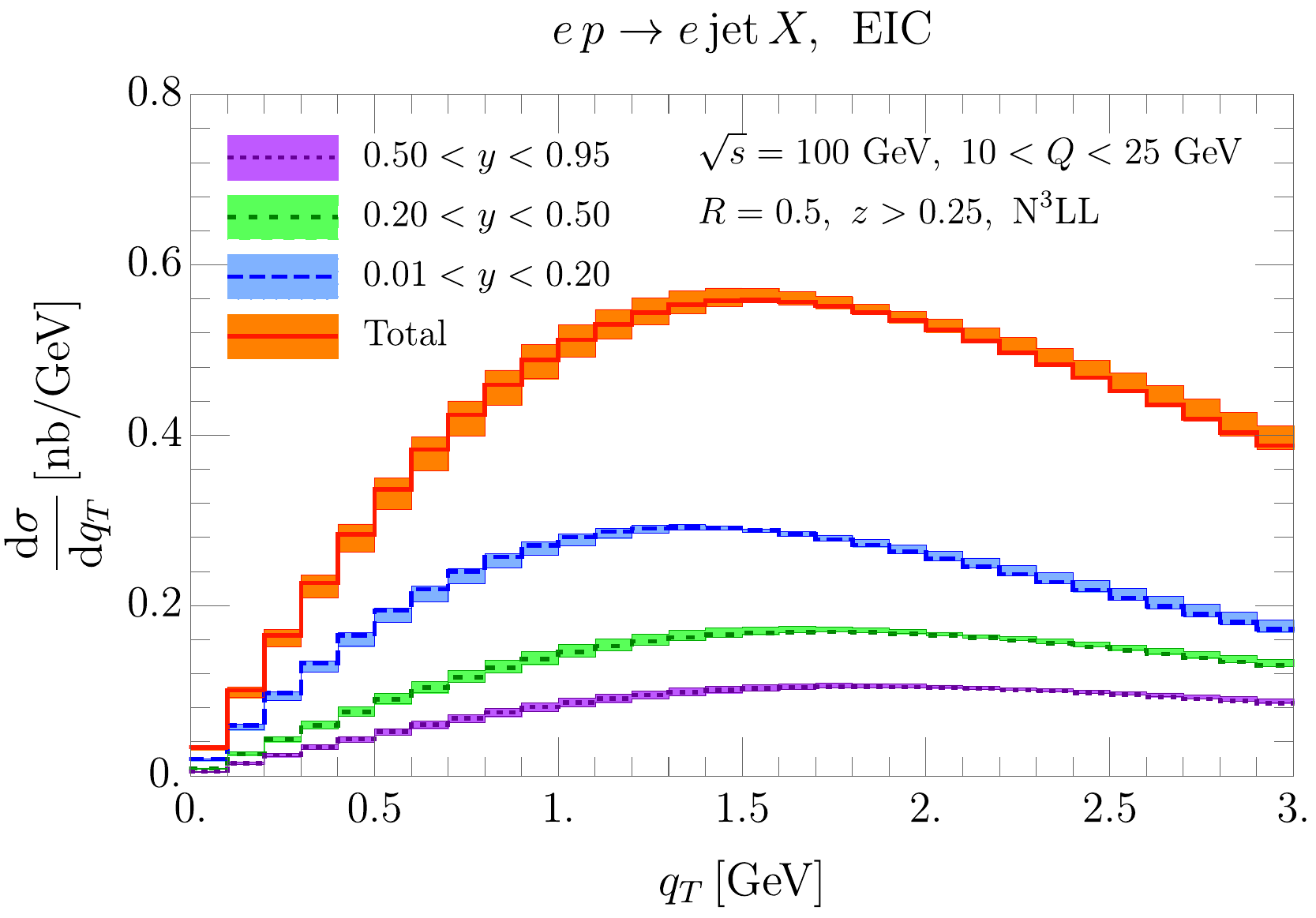}
  \includegraphics[width=0.49\textwidth]{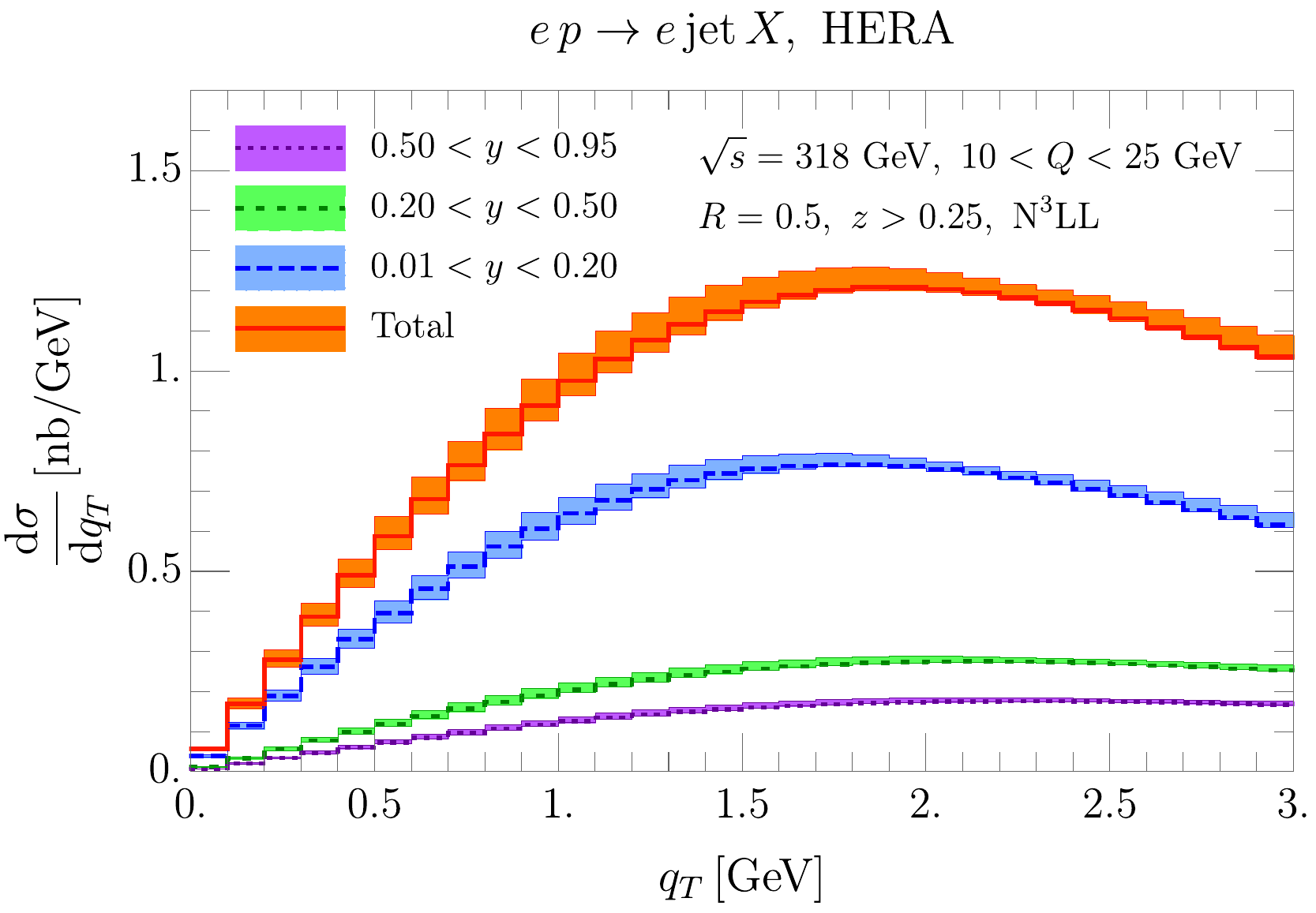}\\
  \caption{TMD cross section for SIDIS with jets at the EIC (left) and at HERA (right), with $10 < Q < 25$ GeV and different intervals in elasticity within the range $0.01 < y < 0.95$.  Results are obtained with the prescription in \eq{combineLargeR}.}
\label{fig:SIDISvaryY}
\end{figure*}
%%%
Our results are shown in \fig{SIDISvaryY}, for which we consider different intervals in the elasticity $y$ in the range $0.01 < y < 0.95$. In each case, we obtained the uncertainty band by independently varying the scales $\mu_H$ and $\mu_0$ up and down by a factor of $2$ around their central values, and taking the envelope. We find that roughly half of the contribution to the cross section comes from low elasticity ($y<0.2$). The variation in shape between the different elasticity intervals is modest; at high elasticity the peak of the distribution is shifted towards larger transverse momenta.

%%%
\begin{figure*}[tb]
  \centering
  \includegraphics[width=0.49\textwidth]{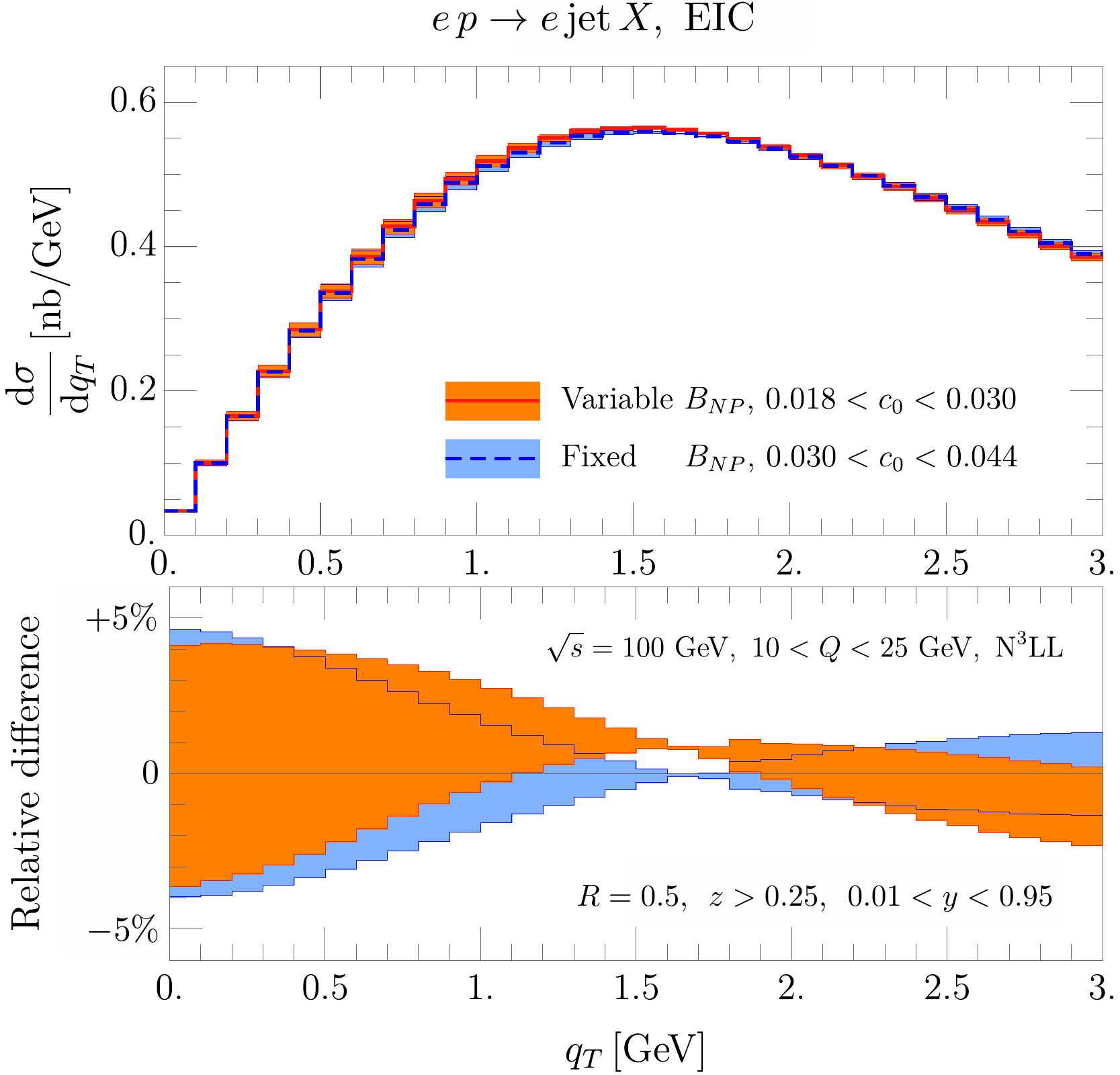}
  \includegraphics[width=0.49\textwidth]{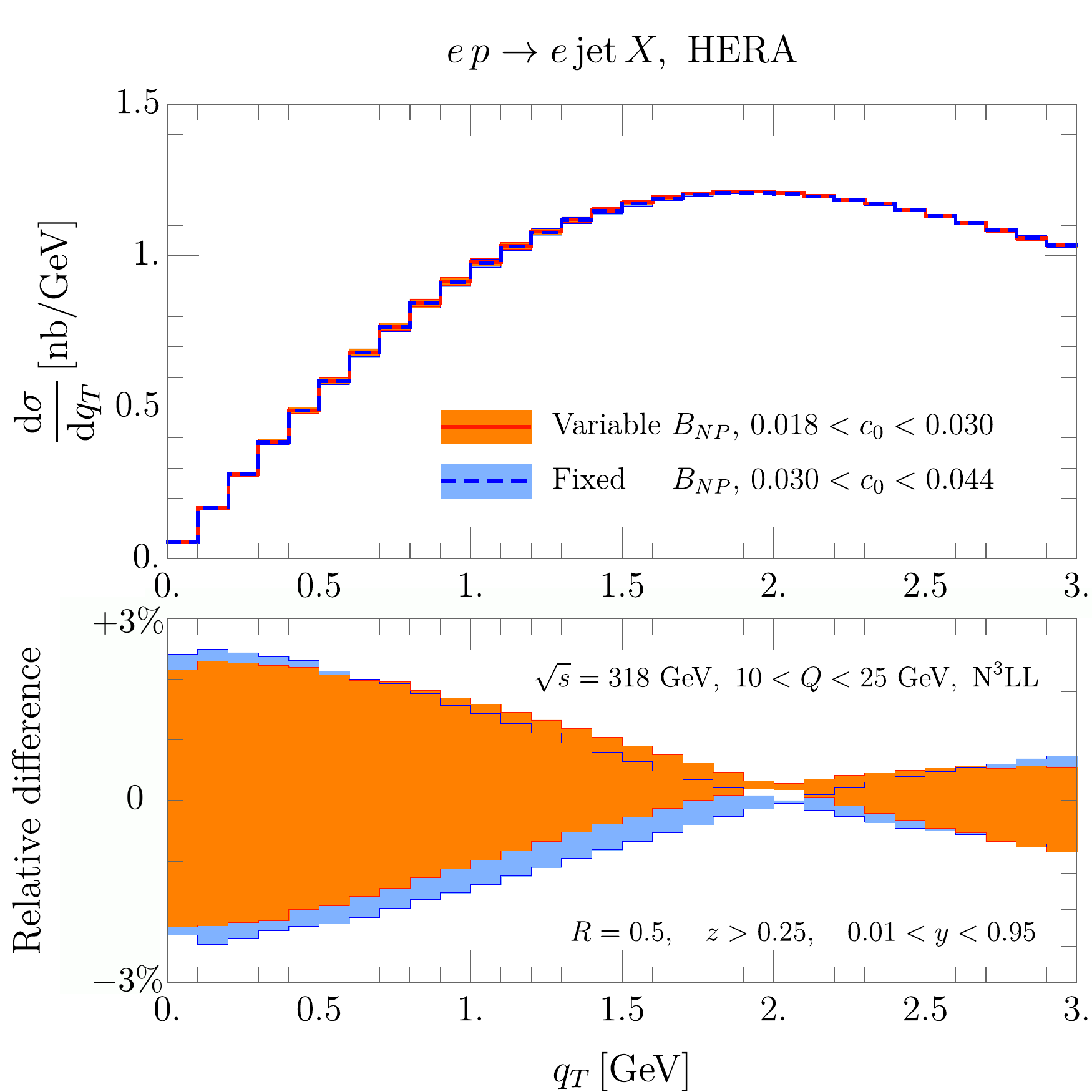}\\	
  \caption{Sensitivity of the cross section to nonperturbative effects at the EIC (left) and HERA (right). This is estimated by varying the parameter $c_0$, that controls the nonperturbative contribution to the evolution kernel, within its current statistical uncertainty~\cite{Bertone:2019nxa}. Results are obtained with the prescription in \eq{combineLargeR}.}
\label{fig:SIDISvaryNP}
\end{figure*}
%%%

We now investigate the sensitivity of our observable to nonperturbative hadronic physics. 
A rough impression can be obtained by varying the parameters $B_{NP}, c_0$ and $\lambda_i$ (see eqs.~\eqref{model:rad}, \eqref{eq:c0def} and \eqref{eq:f_NP}) that enter our nonperturbative model. In principle, these parameters are highly correlated and a full error estimate would require taking data with a large number of replicas, along the lines of the original analysis in ref.~\cite{Bertone:2019nxa}. In practice, we observe that the nonperturbative uncertainty is dominated by the variation of the single parameter $c_0$. Therefore, we obtain a realistic estimate of the size of NP effects by simply varying $c_0$ within its statistical uncertainty, which we show in \fig{SIDISvaryNP}. The effect of varying $c_0$ is not large (below $5\%$ at the EIC and $3\%$ at HERA), but non-negligible, and grows for small $q_T$. This plot suggests that such a measurement can likely be used to improve our knowledge of the nonperturbative part of the evolution kernel, parametrized by $c_0$, which is very relevant because it is universal.
We have explored the dependence on $R$, $z_\cut$ and the range in $Q$ and $y$, finding similar sensitivity to nonperturbative effects.

%%%%%%%%%%%%%%%%%%%%%%%%%%%%%%%%%%%%%%%%%%%%%%%%%%%%%%%%%%%%%%%%%%%%%%%%%%%%%%%%
\section{Conclusions}
\label{sec:conc}
%%%%%%%%%%%%%%%%%%%%%%%%%%%%%%%%%%%%%%%%%%%%%%%%%%%%%%%%%%%%%%%%%%%%%%%%%%%%%%%%

The study of the transverse momentum distribution of the proton can benefit from using jets (instead of hadrons) as final state. A clear advantage is that the jet momentum can be calculated in perturbation theory, while the fragmentation of hadrons is an intrinsically nonperturbative process. We provided an initial formulation of this idea, using a modern definition of jets, in ref.~\cite{Gutierrez-Reyes:2018qez}. There we observed, for the first time, that the cross section for dijet production in $e^+e^-$ collisions and SIDIS with a jet in the final state can have the same factorization as for hadronic TMD measurements, simply replacing a TMD fragmentation function by our TMD jet function. This factorization depends on the jet radius $R$ and recombination scheme, holding only for all values of $R$ if the Winner-Take-All axis is used. In particular,  in the regime of small $q_T$, which is interesting for extracting the intrinsic transverse momentum of partons in the proton, the cross section for the standard jet axis does \emph{not} satisfy the usual TMD factorization.

To explore the ramifications of these ideas, we presented numerical results in this paper for Belle II  and LEP ($e^+e^-$ collisions), and HERA and the EIC (SIDIS), building on the existing \Artemide code. We reported the details of the NLO calculations of the TMD jet function, and have also numerically evaluated the  NNLO contribution in the large-radius limit with \Eventtwo. This was motivated by the observation that the NLO result is well described using the large-$R$ jet function, for all experimental cases we consider. Consequently we can achieve the same N$^3$LL accuracy as in the corresponding hadronic TMD cases. 

We have verified the perturbative convergence of our numerical predictions, achieving perturbative uncertainties of order 5\% in the peak of the distribution at N$^3$LL. We also find that our cross sections have similar sensitivity to nonperturbative effects as the corresponding hadronic case, without the burden of additional nonperturbative effects from fragmentation. Specifically, we have investigated how the cross section changes when varying the nonperturbative parameters within the errors provided in ref.~\cite{Bertone:2019nxa}, concluding that in principle these experiments can provide important constraints on these parameters.
Here we benefit from using the $\zeta$-prescription, which ensures that the nonperturbative parts of the evolution kernel and the rest of the TMD are uncorrelated. 

The nonperturbative effects to the jet TMD have not been estimated in this work. However our factorization theorems ensure that these effects can be included in the definition of the jet functions and are therefore universal, i.e.~the same in $e^+e^-$ collisions and SIDIS. In this respect, the hadronization of jets can be treated in the same way as the nonperturbative part of a hadron TMD, and is therefore expected to be subdominant compared to the nonperturbative part of the evolution. Consequently, jet measurements may provide one of the best ways to constrain the nonperturbative part of the evolution kernel. To reduce the sensitivity to hadronization effects one can consider grooming, which will be investigated in a forthcoming publication~\cite{grooming}.

%%%%%%%%%%%%%%%%%%%%%%%%%%%%%%%%%%%%%%%%%%%%%%%%%%%%%%%%%%%%%%%%%%%%%%%%%%%%%%%%
\begin{acknowledgments}
We thank Alexey Vladimirov for assistance with the \Artemide program and feedback on this manuscript. D.G.R.~and I.S.~are supported by the Spanish MECD grant FPA2016-75654-C2-2-P. D.G.R.~acknowledges the support of the Universidad Complutense de Madrid through the predoctoral grant CT17/17-CT18/17. W.W.~and L.Z.~are supported by ERC grant ERC-STG-2015-677323. W.W.~also acknowledges support by the D-ITP consortium, a program of the Netherlands Organization for Scientific Research (NWO) that is funded by the Dutch Ministry of Education, Culture and Science (OCW). This article is based upon work from COST Action CA16201 PARTICLEFACE, supported by COST (European Cooperation in Science and Technology).
\end{acknowledgments}
%%%%%%%%%%%%%%%%%%%%%%%%%%%%%%%%%%%%%%%%%%%%%%%%%%%%%%%%%%%%%%%%%%%%%%%%%%%%%%%%

\appendix

%%%%%%%%%%%%%%%%%%%%%%%%%%%%%%%%%%%%%%%%%%%%%%%%%%%%%%%%%%%%%%%%%%%%%%%%%%%%%%%%
\section{Perturbative ingredients}
\label{sec:pert}
%%%%%%%%%%%%%%%%%%%%%%%%%%%%%%%%%%%%%%%%%%%%%%%%%%%%%%%%%%%%%%%%%%%%%%%%%%%%%%%%

%===============================================================================
\subsection{Conventions}
\label{sec:conventions}
%===============================================================================

The one-loop splitting functions are
%%%
\begin{align}
  p_{qq}(z) &= (1+z^2)\cL_0(1-z)
  \,, \qquad
  p_{gq}(z) = \frac{1-(1-z)^2}{z}\,,
\end{align}
%%%
where the plus distribution $\cL_0$ is defined in \eq{def:+distributions}.
We introduce the following shorthand for logarithms 
%%%
\begin{align}\label{eq:def:logs}
  L_R = \ln \Big(\frac{\mu^2}{E^2\cR^2}\Big)\, ,\qquad
  \mathbf{l}_X = \ln \Big( \frac{\mu^2}{X}\Big)\, , \qquad
  L_\mu = \ln \Big(\frac{b_T^2 \mu^2}{4e^{-2\gamma_E}}\Big)\, ,
\end{align}
%%%
where $X \in \{Q^2,s,\zeta\}$. 

%===============================================================================
\subsection{Plus distributions}
\label{sec:plus}
%===============================================================================

We define dimensionless plus distributions as
%%%
\begin{align} \label{eq:def:+distributions}
  \cL_n (x) = \Big[\frac{\ln^n x}{x}\Big]_+\,,
\end{align}
%%%
which satisfy
%%%
\begin{align}
  \int_0^1 \df x\, \cL_n (x) = 0
\,.\end{align}
%%%
Integrating these plus distributions against a smooth function $f(x)$ results in
%%%
\begin{align} \label{eq:+EmpiricalDefinition}
  \int_0^{x_0}\!\! \df x \, f(x) \cL_n (x) = 
  \int_0^{x_0}\!\! \df x \, \frac{\ln^n(x)}{x}\big[f(x)-f(0)\big] - f(0)\! \int_{x_0}^1 \df x\, \frac{\ln^n(x)}{x}\,.
\end{align}
%%%
Plus distributions in terms of the transverse momentum $q_T$ are derived from \eq{def:+distributions},
%%%
\begin{align}
 \cL_n(q_T,q_0) = \frac{1}{q_0^2}\, \cL_n\Big(\frac{q_T^2}{q^2_0}\Big)
\,,\end{align}
%%%
such that 
%%%
\begin{align}
  \int_0^{p_T^2} \!\df q_T^2\, f(q_T^2) \cL_n(q_T,q_0) = 
  \int_0^{p_T^2/q_0^2} \!\!\df x\, f(q_0^2 x) \cL_n (x)\, .
\end{align}
%%%
Related ``cut'' distributions are defined as
%%%
\begin{align}\label{eq:def:cutDistributions}
  \cL_n^\cut (q_T,q_0) = \cL_n (q_T,q_0)\thh(q_0-q_T)\, .
\end{align}
%%%
Plus distributions naturally arise in the expansion of logarithmically-singular terms in dimensional regularization,
%%%
\begin{align}
  \frac1{(1-z)^{1+\ve}} &= -\frac{1}{\ve}\de(1-z)+\cL_0(1-z)+\cO(\ve)\,,
  \nn \\  
  \frac{\mu^{2\ve}}{q_T^{2+2\ve}} &= -\frac{1}{\ve}\de(q_T^2)+\LzeroQT+\cO(\ve)\, .
\end{align}
%%%

In the calculation of the jet function in momentum space, one encounters terms where the above expansion cannot be used because the divergences in the limits $\ve \to 0$ and $\delta^-_E \to 0$ are mixed by a step function. In particular, expanding \eq{EsingleUnexpanded} involves the identity
%%%
\begin{align} \label{eq:two-dim_distribution}
&\frac{\mu^{2\ve}}{q_T^{2+2\ve}}\,\frac{1-z}{(1-z)^2+(\de^-_E)^2}\,\theta\Big(z-1+\frac{q_T}{E\cR}\Big)  
\nn \\ & \quad
  =
  \de(q_T^2)\bigg\{\de(1-z)\Big[-\frac{1}{2\ve^2}+\frac{1}{\ve}\Big(\ln\de^-_E -\frac{1}{2}L_R\Big)-\frac{1}{4}L_R^2\Big]+
  L_R \cL_0(1-z)-2 \cL_1(1-z)\bigg\}
  \nn \\ & \qquad
  +\LzeroQT \Big[-\de(1-z)\ln\de^-_E +\cL_0(1-z)\Big] - \LzeroQTZ \cL_0(1-z)
\,,\end{align}
%%%
where the last term involves a genuine two-dimensional distribution. This identity was obtained by switching to cumulative distributions in both variables, then expanding in $\delta^-_E$, and finally expanding in $\ve$, 
%%%
\begin{align}
  &\int_0^{p^2}\!\!\df q_T^2\,\int_y^1\df z\,  \frac{\mu^{2\ve}}{q_T^{2+2\ve}}\,\frac{1-z}{(1-z)^2+(\de_E^-)^2}\,\theta\Big(z-1+\frac{q_T}{E\cR}\Big)
   \\ & \quad =
  -\frac{1}{2\ve^2}+\frac{1}{\ve}\Big[\ln\de^-_E - \ln\Big(\frac{\mu}{E\cR}\Big)\Big]
  -2\ln\de^-_E\,\ln\frac{p}{\mu}
  -\ln^2\Big(\frac{\mu}{E\cR}\Big) 
  +\frac{1}{2}\ln\Big(\frac{\mu}{E\cR}\Big)\ln(1-y)
  \nn\\ & \qquad
  +2\ln\frac{p}{\mu}\,\ln(1-y)
  -\ln^2(1-y)+\thh\big({E\cR(1-y)-p\big)}\ln^2\Big(\frac{E\cR(1-y)}{p}\Big)+\cO(\ve,\de_E^-)\, .
\nn\end{align}
%%%
Every term can now be identified as the cumulative of a distribution, resulting in \eq{two-dim_distribution}. We note that the last term in the expansion, defined by
%%%
\begin{align}
  &\int_y^1\!\!\df z\int_0^{p^2}\!\!\df q_T^2\, \LzeroQTZ \cL_0(1-z) = 
  -\thh\big({E\cR(1-y)-p\big)}\ln^2\Big(\frac{E\cR(1-y)}{p}\Big)\, ,
\end{align}
%%%
is only well defined by the prescription in \eq{+EmpiricalDefinition} if one first carries out the integral over $q_T^2$ before the integral over $z$. 

%===============================================================================
\subsection{Anomalous dimensions}
\label{sec:anom_dim}
%===============================================================================

%%%
\begin{table}
   \centering
   \begin{tabular}{l || c | c | c | c }
   \hline \hline
      Order & F.O. & $\Gamma_{\rm cusp}$  & $\gamma_V$ & $\mathcal{D}$ \\
      \hline \hline
      N$^3$LL &$a_s^2$ & $a_s^4$ & $a_s^3$ & $a_s^3$ \\
      \hline
      NNLL & $a_s^1$ & $a_s^3$ & $a_s^2$ & $a_s^2$ \\
      \hline
      NLL & $a_s^0$ & $a_s^2$ & $a_s^1$ & $a_s^1$ \\
     \hline \hline
   \end{tabular}
   \caption{Various orders in resummed perturbation theory, and the fixed-order (F.O.) and resummation ingredients they involve. 
   The fixed-order ingredients are the perturbative expansion of the hard function, jet function and the coefficients in the matching of the TMD PDFs onto collinear PDFs. We also use the PDFs extracted at this order as well, and use the corresponding running of the coupling.}
   \label{tab:ordersLog}
\end{table}
%%%
We now list the anomalous dimensions that enter the double-scale evolution described in \sec{resum}. Our predictions use N$^3$LL resummation by default, corresponding to the first row in table~\ref{tab:ordersLog}. An exception is \fig{convergencePlot}, where we compare different orders to test the convergence of resummed perturbation theory.
We only need the anomalous dimensions for quarks, which we expand as
%%%
\begin{align}
  \Gamma_q^{\rm cusp} = \sum_{n=0}^\infty a_s^{n+1}\Gamma_{q}^{[n]}\,,\qquad
  \gamma_{V,q} = \sum_{n=0}^\infty a_s^{n+1}\gamma_{V,q}^{[n]}\,,\qquad
  \mathcal{D}_q = \sum_{n=1}^\infty a_s^{n}\mathcal{D}_{q}^{[n]}\, .
\end{align}
%%%
The coefficients in the expansion of the cusp anomalous dimension are given by
%%%
\begin{align} \label{eq:GammaCusp}
  \Gamma_q^{\zero} &= 4C_F\,, \nn\\
  \Gamma_q^{\one} &= C_F\bigg[C_A\Big(\frac{268}{9}-\frac{4\pi^2}{3}\Big)
  -\frac{80}{9}n_fT_F\bigg]\,, \nn\\
   \Gamma_q^{\two}&=C_F\bigg[C_A^2\Big(\frac{490}{3}-\frac{536\pi^2}{27}+\frac{88}{3}\zeta_3
  +\frac{44\pi^4}{45}\Big) + C_An_fT_F\Big(-\frac{1672}{27}+\frac{160\pi^2}{27}-\frac{224}{3}\zeta_3\Big)\nn\\
   &+ C_Fn_FT_F\Big(-\frac{220}{3}+64\zeta_3\Big) - \frac{64}{27}\big(n_fT_F)^2\bigg] \,,\nn\\
   \Gamma_q^{[3]} &= 20702 - 5171.9\, n_f + 195.5772\,n_f^2 + 3.272344\, n_f^3\, .
\end{align}
%%%
The fourth-order result was computed numerically in ref.~\cite{Vogt:2018miu}, which also provides a full decomposition in terms of color structures. The non-cusp anomalous dimension is given by
%%%
\begin{align}
  \gamma_{V,q}^\zero &= -6C_F\,, \nn\\
  \gamma_{V,q}^\one &= C_F\bigg[C_F\Big(-3+4\pi^2-48\zeta_3\Big) +
  C_A\Big(-\frac{961}{27}-\frac{11\pi^2}{3}+52\zeta_3\Big) +
  n_fT_F\Big(\frac{260}{27}+\frac{4\pi^2}{3}\Big)\bigg]\,, \nn \\
  \gamma_{V,q}^\two &= C_F\bigg[
  C_F^2\Big(-29-6\pi^2-136\zeta_3-\frac{16\pi^4}{5}+\frac{32\pi^2}{3}\zeta_3+480\zeta_5\Big)\nn\\
  &+ C_FC_A\Big(-\frac{151}{2}+\frac{410\pi^2}{9}-\frac{1688}{3}\zeta_3+\frac{494\pi^4}{135}-
  \frac{16\pi^2}{3}\zeta_3-240\zeta_5\Big)\nn\\
  &+C_Fn_fT_F\Big(\frac{5906}{27}-\frac{52\pi^2}{9}+\frac{1024}{9}\zeta_3-\frac{56\pi^4}{27}\Big)
  + (n_fT_F)^2\Big(\frac{19336}{729}-\frac{80\pi^2}{27}-\frac{64}{27}\zeta_3\Big)\nn\\
  &+ C_A^2\Big(-\frac{139345}{1458}-\frac{7163\pi^2}{243}+\frac{7052}{9}\zeta_3-\frac{83\pi^4}{45}
  -\frac{88\pi^2}{9}\zeta_3-272\zeta_5\Big)\nn\\
  & + C_A n_fT_F\Big(-\frac{34636}{729}+\frac{5188\pi^2}{243}-\frac{3856}{27}\zeta_3+\frac{44\pi^4}{45}\Big)
  \bigg]\,.
\end{align}
%%%
The rapidity anomalous dimension can be conveniently expressed in terms of $\Gamma_q^{\rm{cusp}}$ in \eq{GammaCusp} as 
%%%
\begin{align} \label{eq:DAnomDim}
  \mathcal{D}_q^\one &= \frac{\Gamma_q^\zero}{2}L_\mu\,, \nn \\
  \mathcal{D}_q^\two &= \frac{\Gamma_q^\zero\beta_0}{4}L_\mu^2 +
  \frac{\Gamma_q^\one}{2}L_\mu+\mathcal{D}_q^\two(0)\,, \nn \\
  \mathcal{D}_q^\three &= \frac{\Gamma_q^\zero\beta_0^2}{6}L_\mu^3+
  \Big(\frac{1}{2}\Gamma_q^\one\beta_0 + \frac{1}{4}\Gamma_q^\zero\beta_1\Big)L_\mu^2
  + \Big(2\beta_0\mathcal{D}_q^\one(0)+\frac{1}{2}\Gamma_q^\two\Big)L_\mu + \mathcal{D}_q^\three(0)\,.
\end{align}
%%%
The first two coefficients of the QCD beta function, that enter here, are given by
%%%
\begin{align}
  \beta_0 &= \frac{11}{3}C_A - \frac{4}{3}n_fT_F\,, \nn \\
  \beta_1 &= \frac{34}{3}C_A^2-\frac{20}{3}C_An_fT_F-4C_Fn_fT_F\, ,
\end{align}
%%%
and the constant terms read
%%%
\begin{align}
  \mathcal{D}_q^\two(0) &= C_F C_A \Big(\frac{404}{27}-14\zeta_3\Big)-\frac{112}{27}C_Fn_fT_F \,, \nn \\
  \mathcal{D}_q^\three(0) &= C_F\bigg[C_A^2\Big(\frac{297029}{1458}-\frac{1598\pi^2}{243}
  -\frac{6164}{27}\zeta_3-\frac{77\pi^4}{270}+\frac{44\pi^2}{9}\zeta_3+96\zeta_5\Big)\nn\\ \nn
  &+C_An_fT_F\Big(-\frac{62626}{729}+\frac{412\pi^2}{243}+\frac{904}{27}\zeta_3-\frac{2\pi^4}{27}\Big)
  +(n_fT_F)^2\Big(\frac{3712}{729}+\frac{64}{9}\zeta_3\Big)\\
   &+C_Fn_fT_F\Big(-\frac{1711}{27}+\frac{304}{9}\zeta_3+\frac{8\pi^4}{45}\Big) \bigg]\, .
\end{align}
%%%

%===============================================================================
\subsection{Fixed-order ingredients}
\label{sec:fixed_order}
%===============================================================================

The hard function for electron-positron annihilation up to two loop is \cite{Kramer:1986sg,Matsuura:1988sm,Gehrmann:2010ue}
%%%
\begin{align}\label{eq:He+e-}
  H_{e^+e^-}(s,\mu) & = \nn1+2a_sC_F\bigg(-\textbf{l}^2_s-3\textbf{l}_s-8+\frac{7\pi^2}{6}\bigg)+
  2a_s^2C_F\bigg\{C_F\bigg[\textbf{l}^4_s + 6\textbf{l}^3_s+\Big(25-\frac{7\pi^2}{3}\Big)\textbf{l}_s^2 \\ \nn
  &\qquad\qquad+\Big(\frac{93}{2}-5\pi^2-24\zeta_3\Big)\textbf{l}_s + \frac{511}{8}-\frac{83\pi^2}{6}-30\zeta_3
  +\frac{67\pi^4}{60}\bigg] \\ \nn& \quad +
  C_A\bigg[-\frac{11}{9}\textbf{l}^3_s +\Big(-\frac{233}{18}+\frac{\pi^2}{3}\Big)\textbf{l}_s^2
  +\Big(-\frac{2545}{54}+\frac{22\pi^2}{9}+26\zeta_3\Big)\textbf{l}_s \\ \nn
  &\qquad\qquad-\frac{51157}{648}+\frac{1061\pi^2}{108} + \frac{313}{9}\zeta_3-\frac{4\pi^4}{45}\bigg] 
  \\ & \quad + n_fT_F\bigg[\frac{4}{9}\textbf{l}_s^3+\frac{38}{9}\textbf{l}_s^2+
   \Big(\frac{418}{27}-\frac{8\pi^2}{9}\Big)\textbf{l}_s
   +\frac{4085}{162}-\frac{91\pi^2}{27}+\frac{4}{9}\zeta_3\bigg]\bigg\} +\cO(a_s^3)\,,
\end{align}
%%%
where $\textbf{l}_s$ is defined in \eq{def:logs}.
DIS  is related to $e^+e^-$ at the level of the amplitude by $s \to -Q^2$. For the hard function this leads to
%%%
\begin{align}\label{eq:HDIS}
  H_{\rm DIS}(Q^2,\mu) &= H_{e^+e^-}(Q^2,\mu) - 2a_s\pi^2C_F 
  \nn\\  & \quad
  +2a_s^2\pi^2C_F\bigg[
  C_F\Big(2\textbf{l}_{Q^2}^2+6\textbf{l}_{Q^2}+16-\frac{4}{3}\pi^2\Big) 
  \nn \\ &\quad 
  + C_A\Big(-\frac{11}{3}\textbf{l}_{Q^2}-\frac{233}{18}+\frac{\pi^2}{3}\Big)
  + n_fT_F\Big(\frac{4}{3}\textbf{l}_{Q^2}+\frac{38}{9}\Big)\bigg] + \cO(a_s^3) \, .
\end{align}
%%%
The respective tree-level cross sections are given by
%%%
\begin{align}
 \sigma_{0}^{e^+e^-} (s)&= \sum_q \frac{4\pi\al^2 N_c}{3s}\,\bar{e}_q^2(s),\\
  \sigma_{0,q}^{\rm DIS} (Q^2,x)&= \frac{2\pi\al^2}{Q^4}\bigg[1+\Big(1-\frac{Q^2}{xs}\Big)^2\bigg]\,\bar{e}_q^2(Q^2)\, ,
\end{align}
%%%
where the effective lepton charge $\bar{e}^2_q$ includes the contribution from $Z$ boson production, 
%%%
\begin{align}
  \bar{e}^2_q(Q^2) = e_q^2 + \frac{(v_q^2+a_q^2)(v_\ell^2+a_\ell^2)-2e_qv_qv_\ell(1-m_Z^2/Q^2)}{(1-m_Z^2/Q^2)^2+m_Z^2\Gamma_Z^2/Q^4}\, .
 \end{align}
 %%%
Here $e_q$ is the electric charge of the quark, $v_i$ and $a_i$ are its vector and axial couplings, $m_Z$ is the mass of the $Z$ and $\Gamma_Z$ its decay width. Our numerical predictions always include $Z$ boson corrections, though their effect is small for Belle and SIDIS.

We computed the (renormalized) NLO quark jet functions in section \ref{sec:jet}. The final expression for the renormalized jet function in transverse momentum space is given by
%%%
\begin{align} \label{eq:JetFunction_q}
  J_{q}^{\one,\, \rm{axis}}(z,\vecb{q},E\cR,\mu,\ze) &= 2C_F\bigg\{ \de(1-z)
  \Big[\frac{3}{2}L_R\de(q_T^2)-
  \textbf{l}_\zeta\,\LzeroQT-\LoneQT+d_q^{\rm axis}(q_T^2)\Big] \nonumber \\ \nonumber
  &\quad +\Big(p_{qq}(z)+p_{gq}(z)\Big)\Big[L_R\de(q_T^2)+\LzeroQT-\LzeroQTZ \Big]\\ 
  &\quad -2\Big[\Big(-3+\frac{2}{z}\Big)\ln(1-z)+2\cL_1(1-z)\Big]\de(q_T^2)\bigg\}\, ,
\end{align}
%%%
where the axis-dependent functions are simply related to \eq{axisDependentDeltas} by
 $d_q^{\rm axis} = \Delta_q^{\rm axis}-\frac{\pi^2}{12}$ and  the $\pi^2/12$ difference comes from the soft function in \eq{NLOsoftFunctionExpanded},
%%%%
\begin{align}
  d_q^\SJA (q_T^2)& = \de(q_T^2) \Big(\frac{13}{2}-\frac{3\pi^2}{4}\Big)\, , \nn\\
  d_q^\WTA (q_T^2) & =\de(q_T^2) \Big(\frac{7}{2}-2\ln^2 2 - \frac{5\pi^2}{12}\Big) 
  + \thetaQT\frac{1}{q_T^2}\bigg[\frac{3q_T}{E\cR}+2\ln\Big(1-\frac{q_T}{E\cR}\Big)\bigg]	
  \nn \\ & \quad
 + \Big(2\ln 2-\frac{3}{2}\Big)\LzeroQTR -\LoneQTR\, .
\end{align}
%%%%

In impact-parameter space the renormalized jet function reads
%%%
\begin{align} \label{eq:JetFunction_b}
  J_{q}^{\one,\, \rm{axis}}(z,\vecb{b},E\cR,\mu,\ze) &=2 C_F\biggl\{ \de(1-z)
  \Big[\frac{3}{2}L_R-\frac{1}{2}L_\mu^2+L_\mu \textbf{l}_\zeta+\tilde d_q^{\rm axis}(B_{ER})\nn\\
  &\quad +\Big(p_{qq}(z)+p_{gq}(z)\Big)\Big[L_R-L_\mu-2\ln(1-z)\nn\\
  &\quad +B_{ER}^2(1-z)^2\,_2F_3\left(1,1;2,2,2;-B_{ER}^2(1-z)^2\right)\Big]\biggr\},
\end{align}
%%%
where the axis-dependent functions are again related to \eq{axisdep} by
 $\tilde d_q^{\rm axis} =\tilde \Delta_q^{\rm axis}-\frac{\pi^2}{12}$,
%%%%
\begin{align}
 \tilde d_q^\SJA& = \frac{13}{2}-\frac{3\pi^2}{4}\, , \nn \\
  \tilde d_q^\WTA & = \frac{13}{2}-\frac{3\pi^2}{4}+\mathcal{G}(B_{ER}).
\end{align}
%%%%
The explicit expression for the function $\mathcal{G}$ entering the impact-parameter space calculation is given below,
%%%
\begin{align}  \label{eq:cG}
\mathcal{G}(B_{ER})=&-11-\frac{5}{8} B_{ER}^2 \, _2F_3\left(1,1;2,2,2;-\frac{B_{ER}^2}{4}\right)-2 B_{ER}^2 \, _2F_3\left(1,1;2,2,2;-B_{ER}^2\right)\ln 2\nn\\
&+\left(4 \pi B_{ER}^2 H_0^S(B_{ER})+\frac{3}{2} \pi  H_0^S(B_{ER})-8 B_{ER}\right) J_1(B_{ER})\nn\\
&+\left(-4 \pi  B_{ER}^2 H_1^S(B_{ER})+8 B_{ER}^2-\frac{3}{2} \pi  H_1^S(B_{ER})+11\right) J_0(B_{ER})+\mathcal{S}
\end{align}
%%%
where $H_n^S$ are the Struve functions of order $n$. $\mathcal{S}$ is a remainder that we did not manage to simplify further,
%%%
\begin{align}\label{eq:Sapprox}
\mathcal{S}=2B_{ER}^2\sum_{n=0}^\infty \frac{\Gamma(1+n)}{\Gamma^3(2+n)}(-B_{ER}^2)^n\left[H_{2n}-n\,_3F_2\left(1,1,1-2n;2,2;\frac{1}{2}\right)\right]
\end{align}
%%%
with $H_n$ the $n$-th harmonic number.

In \eq{JetFunction_b} we already absorbed the soft function (and removed the soft-collinear overlap) as described in \sec{resum}, and the expressions are therefore free of divergences. For completeness we list the soft function at NLO~\cite{GarciaEchevarria:2011rb,Echevarria:2015uaa}
%%%
\begin{equation}\label{eq:NLOsoftFunctionExpanded}
  S_q^\one(\vecb{b},\mu,\ze) = -4C_F\bigg[
  \frac{1}{\ve^2}-\frac{1}{\ve}\ln\Big(\frac{\de^+\de^-}{\mu^2}\Big)-\ln\Big(\frac{\de^+\de^-}{\mu^2}\Big)L_\mu-\frac{1}{2}L_\mu^2-
  \frac{\pi^2}{12}\bigg] + \cO(\ve)\, .
\end{equation}
%%%

For $\theta \gg R$, the dependence on the axis vanishes and the jet function factorizes according to \eq{smallR}. The semi-inclusive quark jet function that enters in this expression is at NLO given by~\cite{Kang:2016mcy}
%%%
\begin{align}\label{eq:QsiJF}
  \mathcal{J}^\one_{q}(z,2z E{\cR},\mu) = &2C_F\bigg[
  \de(1-z) \Big(\frac{13}{2}-\frac{2\pi^2}{3}+\frac{3}{2}L_R\Big)+(L_R-2\ln z)\Big(p_{qq}(z)+p_{gq}(z)\Big)\nonumber\\
  &-2p_{gq}(z)\ln(1-z)-2(1+z^2)\cL_1(1-z)-1\bigg],
\end{align}
%%%
The one-loop matching coefficients for TMD fragmentation from quarks are \cite{Collins:2011zzd,Echevarria:2016scs} 
%%%
\begin{align}\label{eq:QTMDmatching}
z^2\mathbb{C}^\one_{q\to q}(z,\vecb{b},\mu) &= 2C_F\Big[
  p_{qq}(z)\big(2\ln z-L_\mu\big)+\de(1-z)\Big(-\frac{1}{2}L_\mu^2+L_\mu\mathbf{l}_\zeta-\frac{\pi^2}{12}\Big)+1-z\Big], \nn \\
  z^2\mathbb{C}^\one_{q \to g}(z,\vecb{b},\mu) &= 2C_F\Big[p_{gq}(z)\big(2\ln z-L_\mu\big) +z\Big]\, .
\end{align}
%%%

\phantomsection
\bibliographystyle{jhep}
\bibliography{dijet_tmd}

\end{document}